\documentclass{aa}

\usepackage{graphicx}
\usepackage{txfonts}
\usepackage[linkcolor = blue]{hyperref} 
\usepackage{xcolor}
\usepackage{placeins}

\usepackage{soul}

\makeatletter
\renewcommand*\aa@pageof{, page \thepage{} of \pageref*{LastPage}}
\makeatother

\usepackage{natbib}
\bibpunct{(}{)}{;}{a}{}{,} 

\begin{document}

   \title{Properties of self-excited pulsations in 3D simulations of AGB stars and red supergiants}


   \author{A. Ahmad,
          B. Freytag
          \and
          S. Höfner
          }

   \institute{Theoretical Astrophysics, Division for Astronomy and Space Physics, Department of Physics and Astronomy, Uppsala University,
                Box 516, 751 20 Uppsala, Sweden\\
                \email{arief.ahmad@physics.uu.se}
         }

   \date{Received 20 July 2022 / Accepted 01 November 2022}


\abstract
   {The characteristic variability of cool giants and supergiants is attributed to a combination of stellar pulsation and large-scale convective flows. Full 3D radiation-hydrodynamical modelling is an essential tool for understanding the nature of these dynamical processes. The parameter space in our 3D model grid of red giants has expanded in recent years. These models can provide many insights on the nature and properties of the pulsations, including the interplay between convection and pulsations.}
   {We treat 3D dynamical models of asymptotic giant branch (AGB) stars and red supergiants (with current masses $1\,M_{\odot} \le M_{\star} \le 12\,M_{\odot}$) similar to observational data. We aim to explore the relation between stellar parameters and the properties of the self-excited pulsations.}
   {Output from global `star-in-a-box' models computed with the \texttt{CO5BOLD} radiation-hydrodynamics code were analysed, particularly in regards to the pulsation properties, to find possible correlations with input and emergent stellar parameters. The fast Fourier transform was applied to spherically averaged mass flows to identify possible radial pulsation periods beneath the photosphere of the modelled stars. Stellar parameters were investigated for correlations with the extracted pulsation periods. }
   {We find that the pulsation periods varied with the stellar parameters in good agreement with the current expectations. The pulsation periods follow Ritter's period-mean density relation well and our AGB models agree with period-luminosity relations derived from observations. A mass estimate formula was derived from the 3D models, relating the stellar mass to the fundamental mode pulsation period and the stellar radius.}
   {While the non-linearity of the interplay between the self-excited pulsations and the self-consistent convection complicates analyses, the resulting correlations are in good agreement with respect to current theoretical and observational understandings.}


   \maketitle
%

\section{Introduction}

Mira variables are located far into the low-temperature side of the Hertzsprung–Russell (H-R) diagram. In this region, Mira (Omicron Ceti) was the first star to have its pulsation period identified and, together with other Mira variables, was amongst the first periodic red variable stars to have been determined \citep[see][for historical remarks]{hoffleitHistoryDiscoveryMira1997}. Mira-type stars are extremely evolved stars at the tip of the asymptotic giant branch (AGB). Low-to-intermediate mass stars (with initial masses in the range $0.8M_{\odot}<M_{\star}<8M_{\odot}$) encounters the AGB phase during their stellar evolution. Variability is common among cool giant stars. Studies \citep[see, for example,][]{kissMultiperiodicitySemiregularVariables1999a, cioniVariabilitySpectralClassification2001, soszynskiOpticalGravitationalLensing2009} have revealed that the differences, in terms of luminosity amplitude and period regularity, are great.

The pulsation periods of AGB stars range from $80$ to $1000$ days. The regularity of the variation of the stellar luminosity over time allows the subdivision of the pulsating variable types within the AGB: Miras, semi-regular (SR), semi-regular variables with regular periodicity (SRa), and semi-regular variables with irregular periodicity, which can be subdivided depending on the presence (SRb), or lack of (SRc), any transient periods being observed  \citep{beckerVariableStarMenagerie1998}. Red supergiants (RSGs) are often classified as SRb and SRc stars, showing visual variability with an ambiguous period or amplitude. The pulsating AGB and RSG stars are collectively called long period variables (LPVs); however, Miras are distinguished by high amplitude luminosity variations ($\Delta m > 2.5$), with a long and singular dominant period, in the low-order radial pulsation mode. Stars belonging to the SR, SRa, SRb, and SRc classes show smaller amplitude variations with multiple pulsation periods. Consequently, these properties have allowed the AGB stars to be identified through observation, in particular by the period-luminosity (P-L) relations. Miras follow a well-constrained linear P-L relation \citep[][]{whitelockAGBVariablesMira2008, whitelockAsymptoticGiantBranch2009} corresponding to the fundamental mode branch, whereas the semi-regular variables have P-L sequences parallel to that of Miras and the positions of the sequences are thought to be dependent on which pulsation mode is dominant \citep[][]{woodMACHOObservationsLMC1999, soszynskiPulsationModesLongperiod2013a, woodPulsationModesMasses2015}.   

Pulsations are not limited to stars that are highly evolved, but are common throughout the H-R diagram. Solar-like pulsations in the range of micro-magnitudes are due to stochastically excited oscillations, whereas Cepheids and RR Lyraes have higher pulsation amplitudes generated by the $\kappa$ mechanism \citep{joshiAsteroseismologyPulsatingStars2015}. However, AGB stars differentiate themselves from such well-known types of pulsating stars in two important respects. First, their visual amplitudes are high. Second, a degree of variability is inherent in the pulsations. Considering these two properties together with their stellar parameters suggests the excitation mechanism of the pulsations is different than stochastic excitation or the $\kappa$ mechanism \citep[see relevant review chapter in][]{catelanPulsatingStars2015}. Modelling considerations for the excitation mechanism are complicated, since many processes interact with each other. The cool red giants have extremely deep convection zones with large convective cells, often introducing surface inhomogeneities. The large convective cells actively being cycled in the interior, through a series of up- and downdrafts, mean there is a complex interplay between convection and pulsations, therefore non-linear models are needed for the high-amplitude pulsations.

Using 1D linear pulsation models, \citet{trabucchiModellingLongperiodVariables2019} were able to reproduce observed periods of overtone modes in evolved giants, but their predictions for the fundamental mode periods do not agree with observed relations. In recent work with non-linear models, however, \citet{trabucchiModellingLongperiodVariables2021a} obtained fundamental mode periods more closely in agreement with observations. This indicates that convection and pulsations are not accurately described in a linear fashion; the interplay needs to be treated non-linearly. Still, the produced models are 1D in nature and the description of convection is parameterised according to the mixing-length theory; which, despite its usefulness, is insufficient to reconstruct accurate stellar interior temperatures and pressure structures \citep[see reviews by][for a thorough discussion]{houdekInteractionConvectionPulsation2015, kupkaModellingStellarConvection2017}. The fine-tuning of the mixing-length parameter to reproduce such structures compliant with the laws of physics risks oversimplifying the physical processes active in stellar interiors. The need to use 3D modelling is exemplified by the fact that convection, turbulence and the excited oscillations are intrinsically 3D in nature. Utilisation of 1D models can complement 3D models in furthering our understanding of pulsations in evolved red giants; they can explore a wider range of stellar parameters and dynamics, being appreciably less computationally expensive than complete 3D models. A further summary on the use of 1D models of convection and pulsations is available in the introduction section of \citet{freytagGlobal3DRadiationhydrodynamics2017}. In conclusion, it is hard to overstate the need for global 3D models in modelling pulsations of evolved red giants. \texttt{CO5BOLD} is a radiation-hydrodynamics code that numerically simulates global stellar convection. In simulations of cool red giants, self-excited pulsations exist as standing pressure waves. The developed surface convection cells are consistent or comparable with surface inhomogeneities seen from observations \citep[][]{freytagThreedimensionalSimulationsAtmosphere2008, freytagGlobal3DRadiationhydrodynamics2017, kravchenkoTomographyCoolGiant2018, paladiniLargeGranulationCells2018, chiavassaOpticalInterferometryGaia2020}. 

In this work, pulsation properties of global 3D models of AGB and RSG stars produced with the \texttt{CO5BOLD} code were extracted and correlations against stellar parameters were investigated. In essence, the results of the simulations were treated as observed data to identify any significant physical trends. The paper is presented in the following manner. The setup of the 3D models, the model properties and how the pulsation properties were extracted are presented in Sect. \ref{sect_mainmethod}. Section \ref{sect_mainresults} outlines our main results, with relevant discussions presented in Sect. \ref{sect_maindiscussions}. Section \ref{sect_mainconclusions} contains our final remarks.

\section{Methods} \label{sect_mainmethod}

In this section, we first give a short overview of the basic properties of the \texttt{CO5BOLD} models that we analyse in this paper. Then we discuss the methods used for extracting pulsation properties. In Sects. \ref{sect_methodradiusdef} and \ref{sect_methodpulsation}, the model \texttt{st26gm07n001} is referenced as the representative standard model, since this model produced reliable pulsations qualitatively similar with most of our models. This model is a $M_{\star}=1 M_{\odot}$ mass AGB model with solar abundances, having the time averaged stellar parameters radius, luminosity, and effective temperature, respectively: $R_{\star}=360\,R_{\odot}$, $L_{\star}=7000\,L_{\odot}$, and $T_{\textrm{eff}}=2700\,\textrm{K}$.

\subsection{Setup of 3D models} \label{sect_modelsstetup}

For our simulations, we used the \texttt{CO5BOLD} code \citep{freytagSimulationsStellarConvection2012a, freytagAdvancesHydrodynamicsSolver2013, freytagBoundaryConditionsCO5BOLD2017}, that numerically integrates the coupled non-linear equations of hydrodynamics and radiation transport. Compressible hydrodynamics equations allow the modelling of travelling pressure waves, standing acoustic modes, and transsonic convective flows in the stellar interior, as well as shocks in the atmosphere. A short-characteristics scheme solves the equations of non-local radiative energy transport, appropriate for the optically thin atmosphere and the optically thick interior. The gas opacities are tabulated and are the result of a merger between tables generated from stellar-atmosphere codes such as PHOENIX \citep{hauschildtParallelImplementationPHOENIX1997} or COMA \citep[see][]{aringerSiOMoleculeAtmospheres2000thesis, aringerSyntheticPhotometryGiants2016} for the outer layers and OPAL data for the stellar interior \citep{iglesiasSpinOrbitInteractionEffects1992}. No radiation pressure on gas is included, so far.

The tabulated equation of state accounts for the ionisation of H and He, the formation of H$_2$ molecules, and a representative neutral metal. Tables that take the frequency dependence of opacities into account are necessary to model the temperature structure of the outer atmospheric layers correctly. That is crucial, for example, for the formation of dust grains \citep{hofnerExploringOriginClumpy2019}. However, most of the models analysed here only use grey (Rosseland mean) opacity tables. These are sufficient for the study of stellar pulsations, and save significant computer time. As the radiation-transport with grey opacities already dominates the overall run time of a simulation, it slows down with frequency-dependent tables by a factor of about two (because of the further reduced radiative time step) times the number of opacity bins employed -- typically five or three for global models \citep[see][respectively]{chiavassaRadiativeHydrodynamicsSimulations2011a, hofnerExploringOriginClumpy2019}.

All models use a 3D cubical computational box and a Cartesian grid, with a grid dimension ranging from 171$^3$ to 765$^3$ cells (see Table~\ref{table:appendixtablemodel}). For most models the computational box comprises the star and has equidistant grid axes in all directions. For a few models, this domain was augmented with an outer box, where the distance between the grid points gradually increases with the distance from the stellar centre. That allows the tracking of dust clouds \citep[see][]{hofnerExploringOriginClumpy2019, hofnerExplainingWindsAGB2022} but is irrelevant for the properties of the pulsating star. The gravitational potential is a prescribed formula assuming that most of the mass is located in a compact, not resolvable stellar core. It follows an $1/r$ law in the outer layers and is smoothed in a core region with radius $r_0$ (typically 20\,\% to 25\,\% of the stellar radius). While several formulas have been tried, the most often used is of the form
\begin{equation}
  \phi(r) = -G*M * (r^4+r_0^4)^{-1/4}
 \enspace .
\end{equation}
Inside a sphere with radius $r_0$, energy is injected at a constant rate according to the desired luminosity.
A drag force is applied that reduces the local velocity in each grid cell in the core region by a small amount per time step, independently of the size of the flow structures. This is in contrast to the action of an artificial viscosity, that is sensitive to velocity gradients. There are options, for instance, to make the drag force stronger in the very centre than in the outer layers of the core. The drag force, acting only in the small core region, damps the dipole flow globally. It also damps smaller-scale convective flows in this core region, where the gravitational potential is smoothed, anyway, and we cannot expect a good representation of the conditions by our simulations. Due to the small amplitude of radial pulsations in the core region, we expect only a small effect of the damping on this type of oscillations. In effect, the drag force leads to a smooth decline of the velocities from the boundary of the core region towards the very centre, with a rate depending on the detailed settings.

The initial models for the first global simulations were produced by integrating 1D hydrostatic stratifications and distributing the data onto a 3D sphere. This led to long transition phases, in which convective structures developed first in the surface layers \citep[see Fig.\,1 in][]{chiavassaRadiativeHydrodynamicsSimulations2011a} and only later extended further down. Such a procedure works well when initial 1D and final 3D stratification are similar, for example, when the stratification is close to radiative equilibrium or nearly adiabatic and the dynamical pressure due to convection and pulsations is negligible. However, in other cases it is preferred to start from an existing well relaxed model. Straight-forward transformations, for instance, the interpolation to a different grid or the expansion to a larger computational box are done via external scripts.
In addition, some control parameters such as stellar mass or core radius can be changed gradually over the first few pulsation cycles during a simulation. During this phase, the density and the internal energy can be adjusted in all grid cells in each time step. That allows the smooth transition from one set of stellar parameters to another, shortening a long transition phase except for the scaling time interval itself. We monitor, for instance, the emerging luminosity, the total energy, entropy, pressure, and temperature in the core region, and the radius of the star over time to ensure our models are relaxed. The initial transition phase is omitted from further analyses.

Our dataset shown in Table~\ref{table:appendixtablemodel} comprises models with current stellar masses of 1 and 1.5 solar masses, representing AGB stars, as well as models between 5 and 12 solar masses, that we take as representative for RSG stars. Earliest \texttt{CO5BOLD} models are omitted, because often the time sequences are rather short for a reliable determination of the pulsation properties, and the simulations were replaced by newer ones with higher numerical resolution spanning longer time intervals. However, we note that the models analysed here are computed with different versions of \texttt{CO5BOLD}, with unequal grid sizes, and other differences in the numerical setup to maximise the available range of stellar parameters.

\subsection{Definition of the stellar radius} \label{sect_methodradiusdef}

In contrast to other stellar parameters, which are more or less explicitly set in the simulations, the stellar radius, $R_{\star}$, is a resulting property of the models. It is important to have a reliable and consistent method for determining the stellar radius of the model. Not only would the pulsation extraction process be affected by an inexact radius, but it would also risk misrepresenting the stellar parameters. To define the stellar radius, one method we used was the same as \citet{freytagGlobal3DRadiationhydrodynamics2017}, who relied on the average of the luminosity and temperature profile of the star, and solved for the radius following the Stefan-Boltzmann law $4\pi\sigma R^{2}T(R)^{4}=L$. Although this definition works well in most cases, it can be unreliable when spherical averages are applied in the presence of significant opacity fluctuations arising due to variations in the surface density and temperature. For example, in models with similar stellar parameters that only differ in whether they are based on grey or non-grey radiative transfer, the radii derived from the latter are calculated to be greater. This is primarily due to fluctuations in the density that naturally occur in such simulations. The fluctuations may introduce sudden opaque structures that can be misinterpreted as the outer boundary of the photosphere which cause the gradient of the temperature profile in the surrounding atmosphere to be less well-defined. Hence definitions that rely on the explicit detection of temperature changes in the atmosphere, including averaging of the temperatures, can lead to inaccurate values of the radius.

Six further values for the radius have been used in this study. Four were defined by setting the requirement of the Rosseland mean opacity ($\tau_\textrm{Ross}$) to equal $2/3$, $1$, or $10$, or at the first instance of a spherically averaged temperature of $T_\textrm{avg}=\textrm{8000}\,\textrm{K}$ going outwards from the centre of the star. Whilst using  $\tau_\textrm{Ross}=2/3 \ \textrm{or} \ 1$ has practical and physical applications in relation to the photospheric radius of a star, the choice of $\tau_\textrm{Ross}=10$ was arbitrary in order to probe regions where it is substantially optically thick, therefore representing an estimate of the inner radius of the star. A similar motivation was behind defining a radius at $T_\textrm{avg}=\textrm{8000}\,\textrm{K}$. The temperature of $8000 \, \textrm{K}$ was chosen since it consistently captures the region where the inner temperature profile experienced the steepest decline going outwards from the centre of the star. This distinct region signified a transition region which accordingly may be taken as the radius of the star.

\begin{figure}[t]
\centering
\includegraphics[width=\hsize]{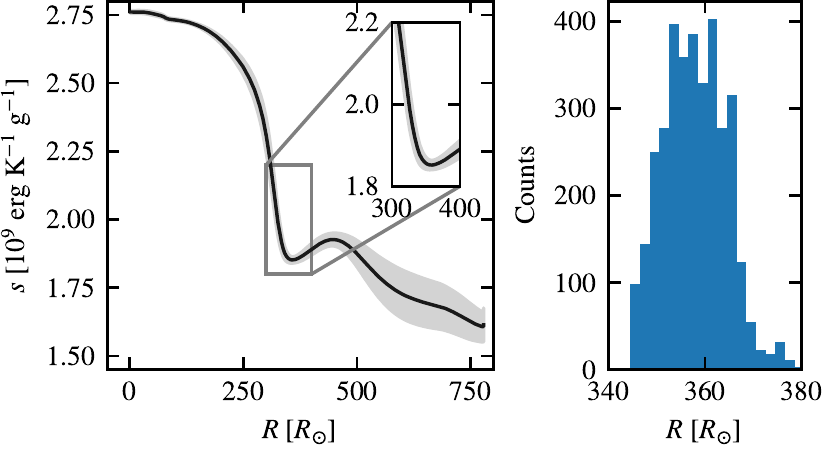}
  \caption{Entropy profile of the standard model \texttt{st26gm07n001}. \textit{Left panel}: Temporal mean value of the entropy as a function of the radial coordinates, with the shaded area being the corresponding standard deviation and the zoomed-in axes shows the first local entropy minimum after the first instance of a spherically averaged temperature of $8000$K away from the centre of the star. \textit{Right panel}: Count values of where the entropy minimum was located over a simulation (in bins of size $2 \, \textrm{R}_\odot$).
          }
     \label{entropydef}
\end{figure}

A new definition is devised in the work presented here, where the radius was determined by the point of the first local minimum of the entropy after a spherically averaged temperature of $T_\textrm{avg}=\textrm{8000}\,\textrm{K}$ from the centre was reached. The entropy minimum marks the thin transition region between a convectively unstable stellar interior and a convectively stable stellar photosphere, and characterises the radius of the star in linear stability analyses and mixing-length theory \citep[see][]{abbettSolarConvectionComparison1997, trampedachImprovementsStellarStructure2014, magicStaggergridGrid3D2015}. For the standard model, the spherically averaged entropy variation across the star is shown in the left panel of Fig.~\ref{entropydef}, where a local minimum can be clearly seen. The location of the entropy minimum depends on the pulsation phase. On the right panel of Fig.~\ref{entropydef}, the radial coordinate of the entropy minimum was plotted as a histogram and illustrates the contraction and expansion phases of the radius, therefore the minimum and maximum radial coordinates respectively, due to pulsations.

%

\begin{figure*}[t]
\resizebox{\hsize}{!}
        {\includegraphics[]{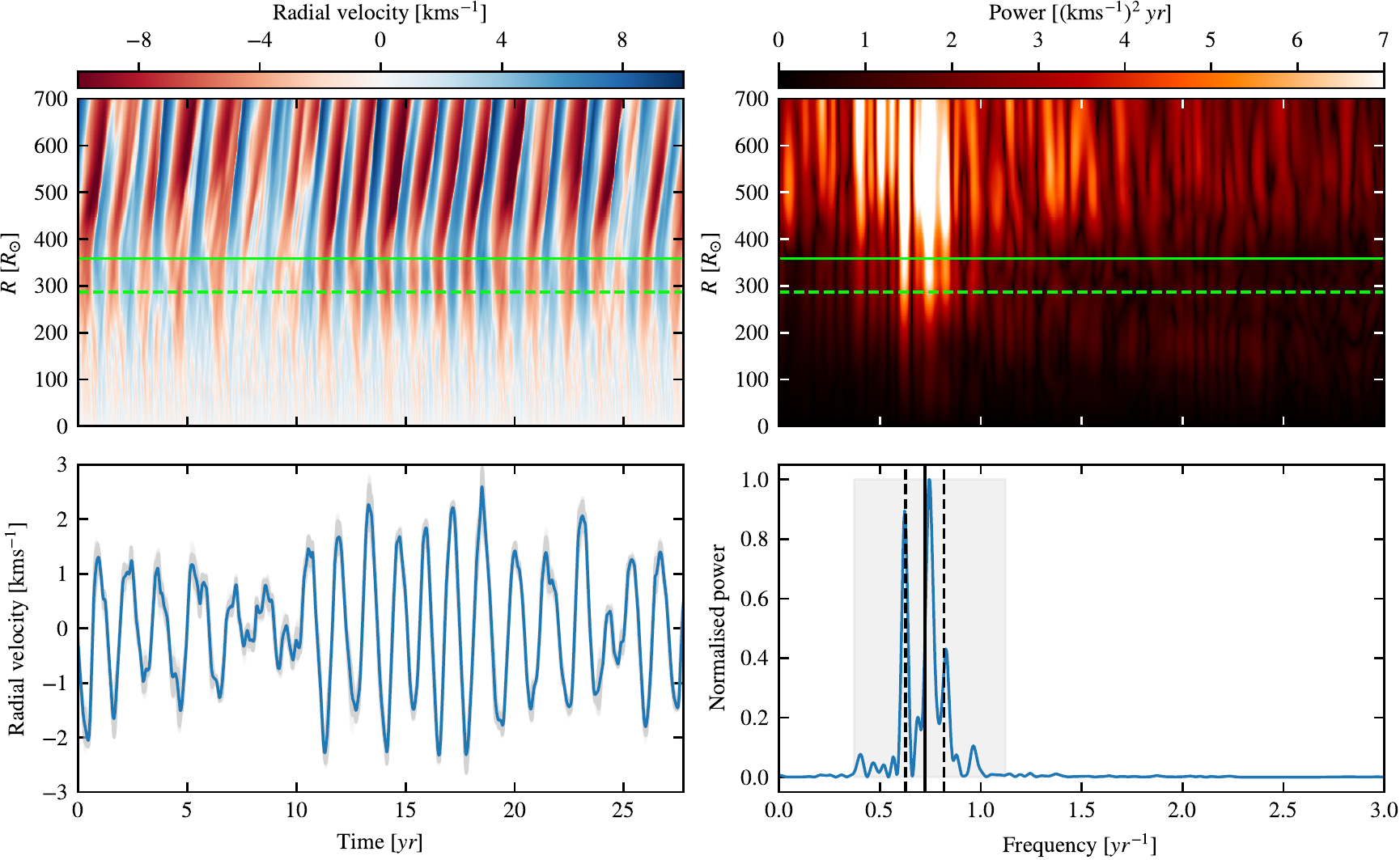}}
          
    \caption{Diagnostics of the spherically averaged radial velocity for the standard model \texttt{st26gm07n001}. \textit{Top left}: Radial velocity as a function of time and radial coordinates, where the solid and dashed lines mark the radius and $80\,\%$ of the radius, respectively. \textit{Bottom left}: Average and standard deviation of the radial velocity (in blue and shaded in grey, respectively) within the layer of $80-100\,\%$ of the radius. \textit{Top right}: FFT power spectrum (values are clipped to the maximum power calculated in between the $80-100\,\%$ of the radius) as a function of the frequency and radial coordinates. \textit{Bottom right}: Averaged and normalised FFT power spectrum across the aforementioned layer. Here the shaded region is the extent over which the weighted averaging (see text) to get a characteristic pulsation frequency was done, while the solid and dashed black vertical lines indicate the computed characteristic frequency and its spread, respectively.}
     \label{Figmainmethod}
\end{figure*}
%

We set an additional definition for the radius for the purpose of capturing the transition region between standing and travelling waves. As will be shown in Sect. \ref{sect_radiuscomparison}, this transition region is helpful in verifying which option for the stellar radius definition is best. Capturing the transition region involved determining the extent to which the pulsations are in phase throughout the layers near (just below, in, and just above) the stellar radius. This determination can be broken down into five steps: 1) the phase of the averaged radial velocity, $\phi_{\textrm{mean}}$, over a thin height across the radius (following the entropy minimum definition) was computed; 2) the phase of the radial velocity of the layers $30\,\%$ below and above the radius  was computed and the phase difference, $\Delta \phi$, relative to $\phi_{\textrm{mean}}$ was calculated; 3) the standard deviation of $\Delta \phi$ across the region was multiplied by $2$ and used as a threshold; 4) new lower and upper boundaries of the radius were determined from the intersection between $\Delta \phi$ and the threshold; 5) averaging was done between the lower and upper boundaries to represent the new radius, determined by the phase. 

In total, we consider seven definitions for the stellar radius; each definition can provide different insights into the models and has its own advantages and disadvantages. As described in detail in Sect. \ref{sect_radiustest}, using the entropy minimum definition provides the best radius definition in the context of analysing the pulsation properties.

\begin{figure*}[t]
\resizebox{\hsize}{!}
        {\includegraphics[]{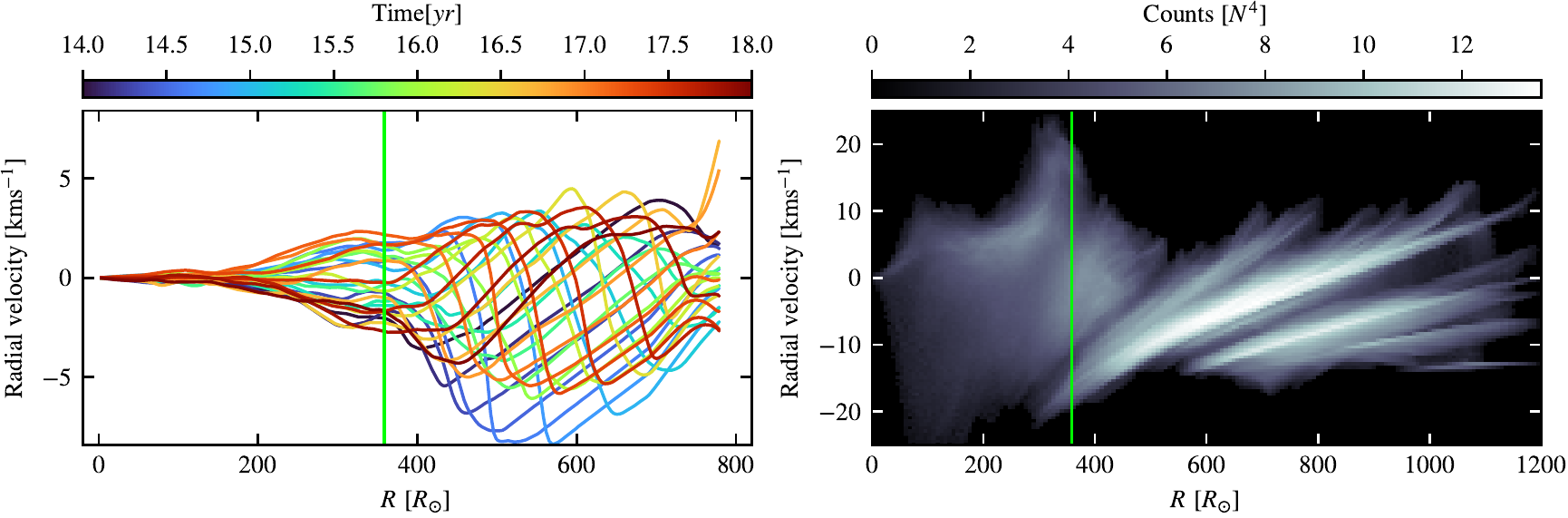}}
          
    \caption{Radial velocity profiles for the standard model \texttt{st26gm07n001}. \textit{Left}: Snapshots of the spherically averaged radial velocity as a function of the radial distance, covering $4$ years of simulated time (about three pulsation cycles). \textit{Right}: Radial velocity computed at each cell of the computational box determined from a snapshot at time $t=14.2 \, [\textrm{yr}]$, presented to show the number of points ($N$) having the radial velocity at a certain distance from the centre of the box. The vertical lines mark the location of the stellar radius.}
     \label{Fig_radialvelocityprofile}
\end{figure*}
%

\subsection{Pulsation period extraction process} \label{sect_methodpulsation}

As outlined in \citet{freytagGlobal3DRadiationhydrodynamics2017}, to solely analyse radial motions, the radial mass flux and mass density were averaged over spherical shells (spherically averaged quantities are denoted with <.>$_{\Omega}$, with $\Omega$ being the solid angle of a sphere). The ratio of the averaged radial mass flux, <$\rho v_{\textrm{radial}}$>$_{\Omega}(r,t)$ and mass density, <$\rho$>$_{\Omega}(r,t)$, is taken as the radial velocity, thus becoming a function of radial distance and time <$v_{\textrm{radial}}$>$(r,t)$. The corresponding spherically averaged radial velocity for our standard model as functions of time and radial coordinate are as shown in the relevant panels of Figs.~\ref{Figmainmethod}~and~\ref{Fig_radialvelocityprofile}. There are three regions of interest in the top left panel of Fig.~\ref{Figmainmethod}: 1) the deep inner core shows irregular radial velocity structures, which arise from fluctuations in the convective flow; 2) around the photospheric layer, periodic variations of inward and outward flows of the radial velocity suggest regular periodic movements of matter due to pulsations; 3) above and beyond the outer atmosphere, where travelling shock waves dominate. The spherically averaged radial velocity under-represents the actual radial velocity amplitudes. Observed motions, derived from high-resolution spectroscopy line profiles, around the photosphere of Mira variables reach about $10 \, \textrm{km/s}$ outwards, and about $15 \, \textrm{km/s}$ inwards \citep[see][]{hinkleInfraredSpectroscopyMira1978, hinkleTimeSeriesInfrared1984}. The radial velocity calculated at each coordinate point for a single snapshot in the computational box, as presented in the right panel of Fig.~\ref{Fig_radialvelocityprofile}, shows close agreement to the amplitude of the observed motions.

The fast Fourier transform (FFT) was applied to the radial velocity, to transform the signal into the frequency domain. Before the FFT, three pre-processing steps were applied to the input signal, which led to the most consistent results for the resulting FFT: 1) subtraction of the temporal average of the signal from the signal itself; 2) tapering with the Tukey window, which decreases the amplitude gradually at the start and end of the signal (thereby helps with dealing with possible discontinuities between the beginning and end of the signal); and 3) zero-padding to increase the length of the input signal by a factor of $10$, (adding zeros at end of the signal produces higher resolution as there are more frequency points in the FFT). All the models presented in this work have sufficiently small sampling interval in the temporal domain to ensure a high FFT sampling rate thus fulfilling the Nyquist criterion safely. 

The power spectrum at each spatial coordinate is shown in Fig.~\ref{Figmainmethod} and the region with the highest intensity can be taken as the first order estimation of the pulsation frequency. To improve on this estimation, a thin region below $R_{\star}$ was selected, and the power spectrum  was averaged over the layers in this region. The chosen region had to be selected strategically: too high above $R_{\star}$, the shocks would dominate the power spectrum and the frequency extracted would not be the pulsation frequency; on the other hand, too low below $R_{\star}$, convective flows may dominate the power spectrum since the pulsation amplitude decreases significantly, therefore obscuring the dominant pulsation frequency. With these effects in mind and after a calibration study, the optimum region to perform the FFT analysis was calculated to be in a layer between $80-100\,\%$ of the radius of the star. The final height-averaged and normalised power spectrum is shown in its corresponding panel in Fig.~\ref{Figmainmethod}, where dominant peaks are observable. 

The ideal power spectrum would be primarily dominated by a single peak but, due to fluctuations to the pulsation amplitude and possible phase changes,  multiple peaks may surround a dominant peak. The peaks within a region surrounding the main peak should be considered as part of the pulsation mode. To account for the smaller peaks, lower and upper bounds of the region to consider were set to $40\,\%$ below and above the frequency of the strongest peak, respectively. Weighted averaging within the width was done to derive the mean frequency of the power spectrum. The weighted mean of the frequency within the interval was computed as \begin{equation}
    f_{\textrm{mean}}=\frac{\sum_{\textrm{i=0}}^{I} P(f_{\textrm{i}}) f_{\textrm{i}}}{\sum_{\textrm{i=0}}^{I} P(f_{\textrm{i}})} \enspace ,
    \label{equation:weighted_mean}
\end{equation} 
where $I$ is the number of all the frequency points within the width (which varies with the time resolution of each model) and $P$ is the corresponding power of the FFT power spectrum at the specific frequency point $f_{i}$. The corresponding weighted root-mean-square (RMS) is then 
\begin{equation}
    f_{\textrm{RMS}}=\left[ \frac{\sum_{\textrm{i=0}}^{I} P(f_{\textrm{i}}) (f_{\textrm{i}}-f_{\textrm{mean}})^{2}}{\sum_{\textrm{i=0}}^{I} P(f_{\textrm{i}})} \right]^{1/2} \enspace .
\label{equation:weighted_rms}
\end{equation}
%
\begin{figure}[t]
\centering
\includegraphics[width=\hsize]{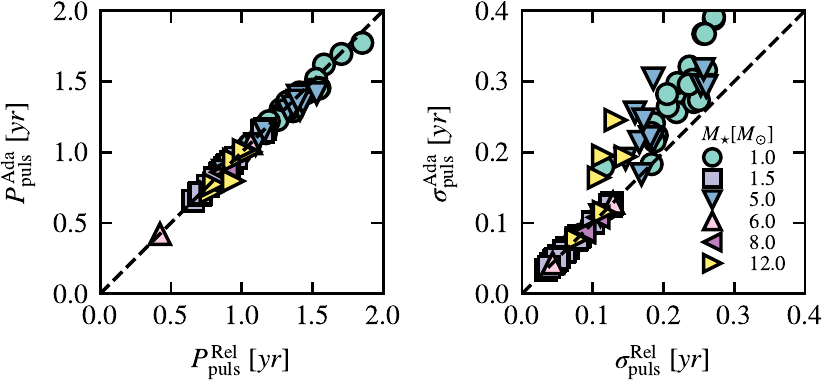}
  \caption{Comparison between the different width methods, relative (Rel) or adaptive (Ada) in superscript, in the derived pulsation periods and spreads (left and right panels, respectively). Two models that have extremely extended atmospheres were omitted from the right panel to preserve clarity of the plot (as the values are around $0.8 \,[\textrm{yr}]$).
          }
     \label{rigrelativeadaptive}
\end{figure}
%
%
\begin{figure*}[t]
\resizebox{\hsize}{!}
        {\includegraphics[]{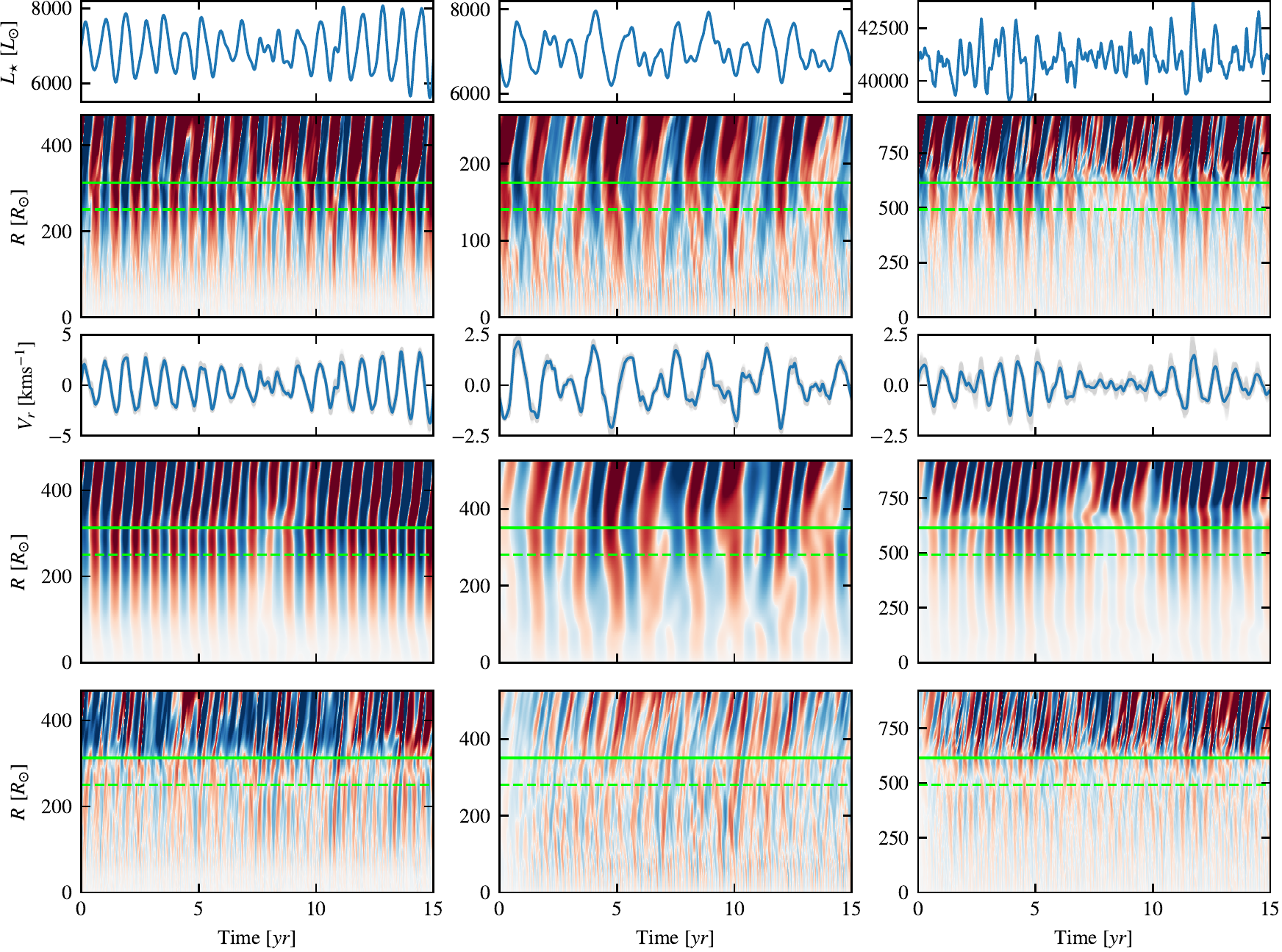}}
  \caption{Pulsation properties of three representative models. Columns from left to right are outputs of selected $1.5 \, M_{\odot}$, $1 \, M_{\odot}$, and $8 \, M_{\odot}$ models (\texttt{st28gm05n023}, \texttt{st28gm06n039}, and \texttt{st34gm02n001}). The rows, from top to bottom (all showing $15$ years of simulated time): bolometric luminosity; spherically averaged radial velocity; its average in a thin layer beneath the stellar radius; recovered velocity field after masking for the pulsation signal; recovered velocity field after removing the pulsation signal. The solid and dashed lines respectively mark the radius and $80\,\%$ of the radius of the star. The limits on the colour scales in the second, fourth, and fifth rows are around $\pm 8$, $\pm 4$, and $\pm 2$ $[\textrm{km/s}]$, respectively, with the colour bars corresponding to the one shown in the top left panel of Fig.~\ref{Figmainmethod}.
          }
     \label{Figcomparetypes}
\end{figure*}

%
The characteristic pulsation frequency and spread were determined from the values $f_{\textrm{mean}}$ and $f_{\textrm{mean}} \pm f_{\textrm{RMS}}$, respectively, and fit into the FFT power spectrum as shown in the vertical lines in the bottom right panel of Fig.~\ref{Figmainmethod}. Due to occasions where a considerable number of smaller peaks were present outside the relative width, an adaptive width was utilised. The adaptive width extends the initial width only if the outer peaks contributed to a change in the final $f_{\textrm{mean}}$ by at least $2\,\%$. This adaptive width did not extend indefinitely as it was also restricted within a range of extension. This is to avoid accounting for higher harmonics due the pulsations or possible overtone modes (where both would occur around twice the initial $f_{\textrm{mean}}$). Thus the adaptive width determination was useful to ensure all or most of the power due to the pulsations was contained within the width used in the weighted averaging. Finally, to convert into the time-domain, the derived pulsation period is simply $P_{\textrm{puls}}=1/f_{\textrm{mean}}$, and its spread $\sigma_{\textrm{puls}}=P_{\textrm{puls}} (f_{\textrm{RMS}}/f_{\textrm{mean}})$. Since there were no nodes between the centre and surface of the star in the radial velocity profiles (evidenced in Figs.~\ref{Figmainmethod}~and~\ref{Fig_radialvelocityprofile}) and no appreciable power was present below the frequency of $f_{\textrm{mean}}$ in the FFT space, we take $P_{\textrm{puls}}$ to be the pulsation period of the fundamental mode.

\section{Results} \label{sect_mainresults}

Given the self-excited nature of the pulsations in the models arising from pressure sound waves due to convection throughout the star \citep[see][for additional background]{freytagGlobal3DRadiationhydrodynamics2017}, the pulsation properties and correlations against various stellar parameters were investigated in this work.

\subsection{Extracted pulsation periods and their spreads}

\begin{figure*}[t]
\resizebox{\hsize}{!}
        {\includegraphics[]{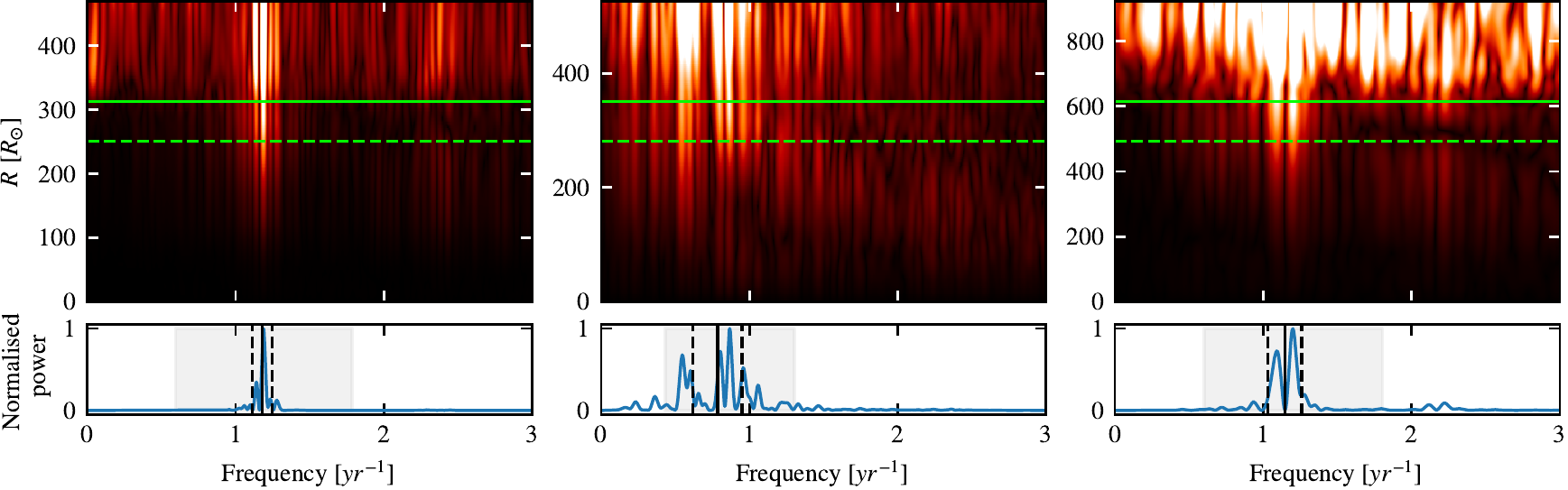}}
  \caption{FFT spectra of the representative $1.5 \, M_{\odot}$, $1 \, M_{\odot}$, and $8 \, M_{\odot}$ models shown in Fig.~\ref{Figcomparetypes} (\texttt{st28gm05n023}, \texttt{st28gm06n039}, and \texttt{st34gm02n001}) respectively. The FFT power spectra are presented as a function of radial coordinates and frequency (scaled to show the maximum power in between $80$-$100\,\%$ of the radius), and with the signal averaged and normalised between the thin layer beneath the stellar radius. 
          }
     \label{Fig_threemodelsfftres}
\end{figure*}

The pulsation periods derived using either the relative or adaptive width method are compared in Fig.~\ref{rigrelativeadaptive}. Clearly, there is little scatter in the derived periods. There is a bigger scatter in the spread, where the adaptive method more consistently results in higher spreads in the pulsation period. This can be expected when the non-linearity of the interplay between pulsations and convection becomes considerable, as there are more peaks surrounding the dominant signal. With more appreciable peaks and the broader width being used in the weighted averaging procedure, larger values on the spread of the pulsation periods are expected when using the adaptive width; which is also suggestive of strong interaction between the pulsations and convection. The overall trend, therefore, of higher divergence from the one-to-one ratio at higher values in the $\sigma_\textrm{puls}$ plot is reasonable. While it is important to keep the $\sigma^\textrm{Ada}_\textrm{puls}-\sigma^\textrm{Rel}_\textrm{puls}$ plot in mind, given the good correlation between the two methods of calculating the $P_\textrm{puls}$, unless otherwise stated, the results presented from now on are using the pulsation properties derived with the relative width method.

\subsection{Representative examples}

This subsection details the information presented in Figs.~\ref{Figcomparetypes}~and~\ref{Fig_threemodelsfftres}, where each column corresponds to a model with a different mass ($1.5 \, M_{\odot}$, $1 \, M_{\odot}$, and $8 \, M_{\odot}$). In Fig.~\ref{Figcomparetypes}, the bolometric luminosity variation and the averaged radial velocity in the layer between $80-100\,\%$ of $R_{\star}$ are presented in the first and third rows, respectively, to indicate the amplitude of the pulsations in the models. 
The bolometric luminosity profiles of the three models show semi-regular variability (of the type SRa). The amplitudes of the $1.5 \, M_{\odot}$ and $1 \, M_{\odot}$ AGB models are low compared to those of typical Miras \citep[see][]{feastInfraredPropertiesMiratype1982,whitelockInfraredColoursMiralike2000}. Disregarding any extinction (circumstellar or interstellar), the two AGB models show apparent bolometric amplitudes of around $\Delta m_{\textrm{bol}}=0.14\, \textrm{mag}$, in contrast to most of the Mira-like pulsations observed in \citet{whitelockInfraredColoursMiralike2000}, which had inferred amplitudes of higher than $\Delta m_{\textrm{bol}}=0.5\, \textrm{mag}$ (after bolometric corrections were applied).

The spherically averaged radial velocity and the corresponding averaged power spectrum after performing the FFT procedure outlined in Sect. \ref{sect_methodpulsation} are shown in the second row of Fig.~\ref{Figcomparetypes} and Fig.~\ref{Fig_threemodelsfftres}, respectively. The fourth row in Fig.~\ref{Figcomparetypes} shows the recovered radial velocity signal in the time-domain, from an inverse FFT after masking was done in the frequency space to isolate only the pulsation frequency. The range of the masking was: $f_\textrm{mean} \pm f_\textrm{RMS}$, which is shown in the bottom row of Fig.~\ref{Fig_threemodelsfftres} (the frequencies within the dashed vertical lines). In contrast, the frequency range that was outside the masked region just described was used to recover the signal that was independent of the primary frequency of the pulsations. This signal is shown in the bottom row of Fig.~\ref{Figcomparetypes}.

In the fourth row of Fig.~\ref{Figcomparetypes}, the pulsations for the $1.5 \, M_{\odot}$ AGB are stable, despite a sudden decrease in the pulsation amplitude just before the 10 year mark. For the less massive $1\, M_{\odot}$ AGB with lower surface gravity, in the middle column, the pulsations are still evident, albeit more irregular. There are fluctuations, phase and amplitude shifts in its pulsations, and this is also reflected in its corresponding FFT power spectrum in Fig.~\ref{Fig_threemodelsfftres} - showing many high amplitude neighbouring peaks around a main signal. For the RSG model in the last column, a strong decrease in the pulsation amplitude can be observed in the middle of the simulation. The amplitude stayed low for about two pulsation cycles, then recovering its previous amplitude. %
\begin{figure}[t]
\centering
\includegraphics[width=\hsize]{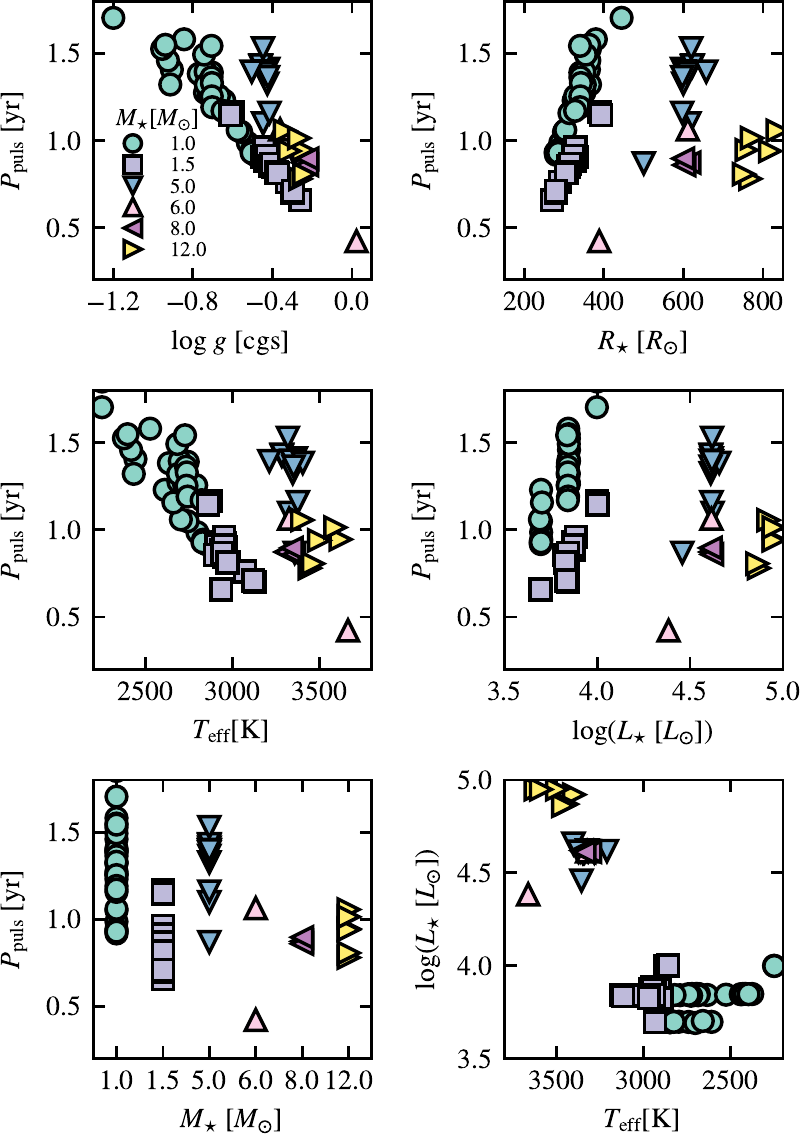}
  \caption{Derived pulsation periods plotted against selected stellar parameters of the models. Five panels show the pulsation period against (top to bottom reading left to right): surface gravity; stellar radius; averaged effective temperature; stellar luminosity and stellar mass. The last panel illustrates how the models in this work fit into the H-R diagram. 
          }
     \label{fig_period_vs_sparams}
\end{figure}
%

The masking to remove the pulsations enables the observation of the radial component of turbulence and convection within the star, shown in the bottom row of Fig.~\ref{Figcomparetypes}. By comparing the relative intensity of the convective signals, the depths of the zone with significant convective energy flux are shown to be different between the models. The density contrast in the core and atmosphere are different, and, as explained in Sect. \ref{sect_modelsstetup}, the inner boundary conditions for different models might vary. Therefore, it is difficult to disentangle and compare any properties of the convective flows in models of different masses. All the representative models show strong overlapping regions where the pulsation and convective signals are strong, indicating that the interplay between pulsations and convection is definitely possible. 

Finally, we note that the FFT power spectrum of the $8\,M_{\odot}$ model shows a detectable signal just below twice the main pulsation frequency. We verified the signal to be due to an overtone mode in the pulsations. Showing a clear node around $80 \, \%$ of $R_{\star}$ in the power spectrum, the model has the strongest imprint of an overtone in this work. A similar but ambiguous signal is present in the power spectrum of the $1.5 \, M_{\odot}$ model. However, we attribute this signal to be due to the first harmonic of the main pulsation frequency due to the non-sinusoidal nature of the radius variations. Shock waves in the atmosphere contribute to this signal above the radius $R_{\star}$. We speculate that, below the radius $R_{\star}$, interference by convection produces a noticeable signal around twice the pulsation frequency. The relatively large pulsation amplitude makes this feasible. Overall, after investigating the models on an individual basis, strong signals suggesting any overtones are rare. With exception to the $8\,M_{\odot}$ model, we favour the interpretation of the second harmonic being the contributor to any detectable signals about twice the pulsation frequency, which were observed in a few of the models.

\subsection{Pulsation properties and correlations}

The extracted pulsation periods of the stars are plotted against their fundamental stellar parameters in Fig.~\ref{fig_period_vs_sparams}. The plots represent an overview of the current models that we have analysed and the range of stellar parameters we were able to cover so far. Two important trends should be pointed out: 1) increasing pulsation period with radius and; 2) decreasing pulsation period with higher surface gravity. The trends are primarily linked to how the stellar parameters affect the dynamical timescale of the pulsations, and will be presented in further detail alongside other considerations in the following sections.

\subsubsection{Ritter's period-mean density relation}
\label{rittersmeandensity}

\begin{figure}[t]
   \centering
   \includegraphics[width=\hsize]{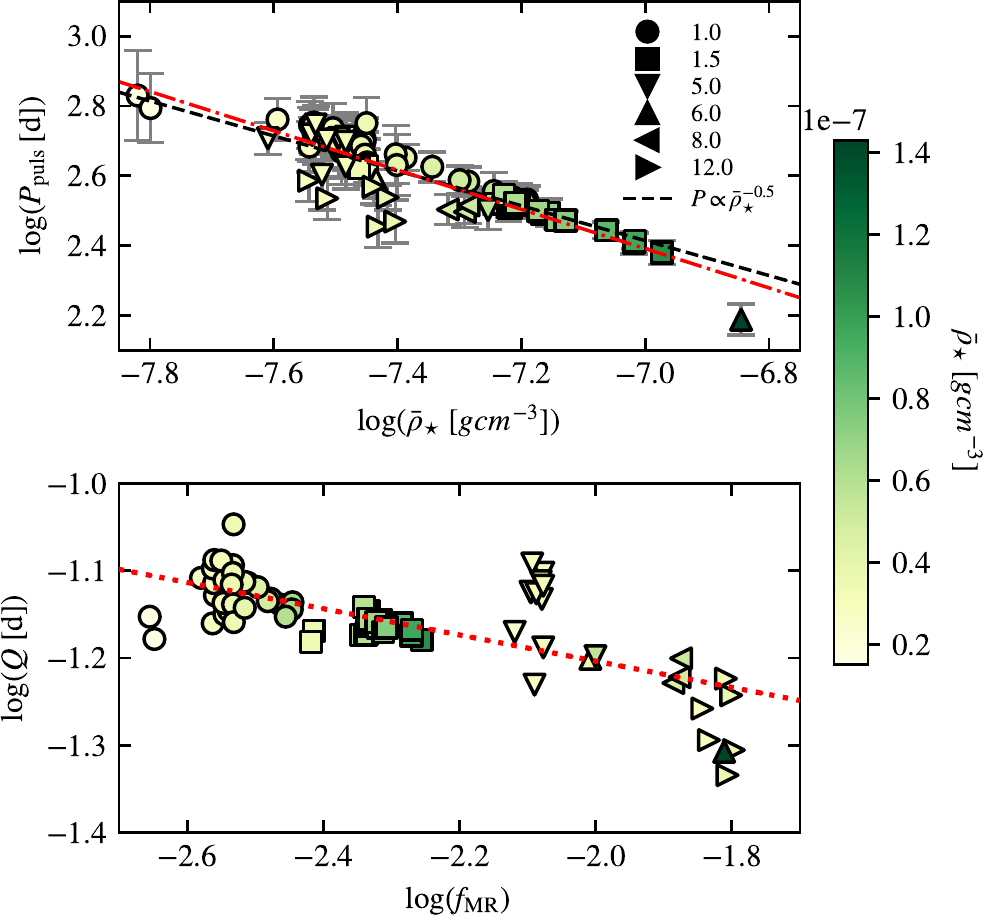}
      \caption{Examining the effect of the stellar mean-density on the pulsation periods. \textit{Top panel}: Extracted fundamental pulsation periods against the mean density. Over-plotted is Ritter's period-mean density relation (dashed line) fitted to the model points and the dash-dotted line is the unconstrained regressed line based on the model points. \textit{Bottom panel}: Pulsation constant, $Q$, against the stellar mass-to-radius ratio, $f_{\textrm{MR}}$, where the dotted line is the result of a linear regression applied to all the models in this plot.
              }
         \label{figrhomr}
   \end{figure}

\citet{ritterUntersuchungenUeberHoehe1879} first established the relationship between the pulsation period and the mean-density of a pulsating homogeneous sphere experiencing adiabatic radial pulsations. The final relation is $P_{\textrm{puls}}\propto \bar{\rho}_{\star} ^{-0.5}$ and in application to stars, the contraction and expansion of the star would be accompanied by a periodic increase and decrease in effective temperature and luminosity. However, the idea of pulsating stars did not gain much attention until \citet{shapleyNatureCauseCepheid1914} suggested radial pulsations, instead of binarity, could explain the variability in Cepheids. \citet{eddingtonProblemCepheidVariables1918, eddingtonPulsationsGaseousStar1919} then provided further proof and a theoretical framework for such pulsating stars and came to the same conclusions as \citet{ritterUntersuchungenUeberHoehe1879}; for radial pulsators where the restoring force is the pressure, the pulsation period of a radial mode can be estimated by the time it takes for a sound wave to travel back and forth from the centre to the surface of the star - defined as the sound-crossing time. In the top panel of Fig.~\ref{figrhomr}, we find a good agreement between our models and Ritter's period-mean density relation. One can derive the pulsation constant, $Q$, from the period mean-density relation, provided in \citet{foxTheoreticalGrowthRates1982}:
\begin{equation}
    P_{\textrm{puls}} = Q \left(\frac{\bar{\rho}_{\star}}{\bar{\rho}_{\odot}}\right) ^{-1/2} \quad 
    \xrightarrow{}
    \quad
    Q=P_{\textrm{puls}} \left( \frac{M_{\star}}{M_{\odot}} \right)^{1/2} \left( \frac{R_{\star}}{R_{\odot}} \right) ^{-3/2} \enspace .
    \label{equation_pulsationconstant}
\end{equation}
Applying a linear regression on the $P_{\textrm{puls}} - \bar{\rho}_{\star}$ plot in Fig.~\ref{figrhomr}, but with ${\bar{\rho}_{\star}}/{\bar{\rho}_{\odot}}$ on the $x$-axis (using ${\bar{\rho}_{\odot}}=1.3963 \,\textrm{gcm}^{-3}$), provides the coefficients:
\begin{equation}
    \begin{array}{l}
        \log(P_\textrm{puls})=-0.5625(\pm0.0331)\log\left({\bar{\rho}_{\star}}/{\bar{\rho}_{\odot}}\right) \ \\ \quad \quad \quad \quad \quad \quad
        -1.6275(\pm0.2487) \enspace .
    \end{array}
    \label{equation:p_rho_fit}
\end{equation}
Therefore, a slightly steeper gradient was obtained against the value derived by Ritter's relation, as can be seen on the top panel of Fig.~\ref{figrhomr}. The intercept of the derived relation gives the fundamental radial pulsation period of the Sun as $0.0236 \pm 0.0182 \, \textrm{days}$, which within its uncertainty agrees with the literature value of $0.033 \, \textrm{days}$ \citep{handlerAsteroseismology2013a}. If a fixed slope of $-0.5$ was used instead, the corresponding intercept would be $0.0695 \pm 0.0111 \,\textrm{days}$, thus significantly overestimating the value of the fundamental radial pulsation period of the Sun. On this premise, the steeper gradient derived is justifiable for the models presented here. However, it is noted that our models follow the $P_{\textrm{puls}} - \bar{\rho}_{\star}^{-0.5}$ relation well, which is noteworthy given that the models do not fit the underlying assumptions of Ritter's relation; derived for a homogeneous and spherically symmetric atmosphere experiencing adiabatic pulsations.

%

Given the dependence of the pulsation constant on the stellar parameters of $Q\propto R_{\star}^{-1}(M_{\star}/R_{\star})^{0.5}$ for a given period, it is useful to derive the relationship between $Q$ and the stellar mass-to-radius ratio, $f_{\textrm{MR}}=(M_{\star}/M_{\odot})/(R_{\star}/R_{\odot})$, as in the bottom panel of Fig.~\ref{figrhomr}. Under the assumption that this relationship is linear and finding the relevant parameters of its relationship through a linear regression, the mass of a pulsating star can be inferred with (by virtue of substituting the relationship into Eq.~(\ref{equation_pulsationconstant})):
\begin{equation}
    M_{\star}=\left(10^{-a} P_\textrm{puls} R_{\star} ^{b-1.5} \right)^{1/(b-0.5)} \enspace ,
    \label{equation:mass_estimate}
\end{equation}
where $P_\textrm{puls}$ being in days and the $M_{\star}$ and $R_{\star}$ in solar units. The parameters $a$ and $b$ derived from the linear regression on $Q$ and $f_{\textrm{MR}}$ are presented in Table~\ref{table:1}, where different values of the parameters were calculated depending on the different type of models being considered in the regression: AGB stars, RSGs or all together. The reasoning behind this is apparent from the bottom panel of Fig.~\ref{figrhomr}, which indicates two separated regions for the AGB stars and RSGs. In addition, despite showing some degree of linear dependence, the R-squared ($R^{2}$) values of the linear regressions on $Q$ and $f_{\textrm{MR}}$ are relatively poor as seen in Table~\ref{table:1}. 
\begin{table}
\caption{Parameters as derived from linear regressions between $Q$ and $f_{\textrm{MR}}$. The parameters $a$ and $b$ are to be substituted into Eq.~(\ref{equation:mass_estimate}) in order to estimate the mass of a star, and $R^2$ is the coefficient of determination from the regressions.}              
\label{table:1}      
\centering                                      
\begin{tabular}{c c c c}          
\hline\hline                        

{} & $a$ & $b$ & $R^{2}$ \\
    AGB  & $-1.5719 \pm 0.1389$                               & $-0.1768 \pm 0.0586$ & $0.57$ \\      
    RSG  & $-2.1250 \pm 0.1730$ & $-0.4737 \pm 0.0885$ & $0.75$ \\
    All  & $-1.5041 \pm 0.0404$ & $-0.1503 \pm 0.0175$ & $0.50$ \\
\hline                                             
\end{tabular}
\end{table}

However, any practical application of Eqs.~(\ref{equation:p_rho_fit}) and (\ref{equation:mass_estimate}) to observed data may be limited. We note that the stellar radius value in the relations represents the entropy minimum radius (see Sect. \ref{sect_methodradiusdef}), which may differ considerably from radius values derived from observations, depending on the optical depth of the visible outer layers.

\subsubsection{P-L relation}

\begin{figure}[t!]
   \centering
   \includegraphics[width=\hsize]{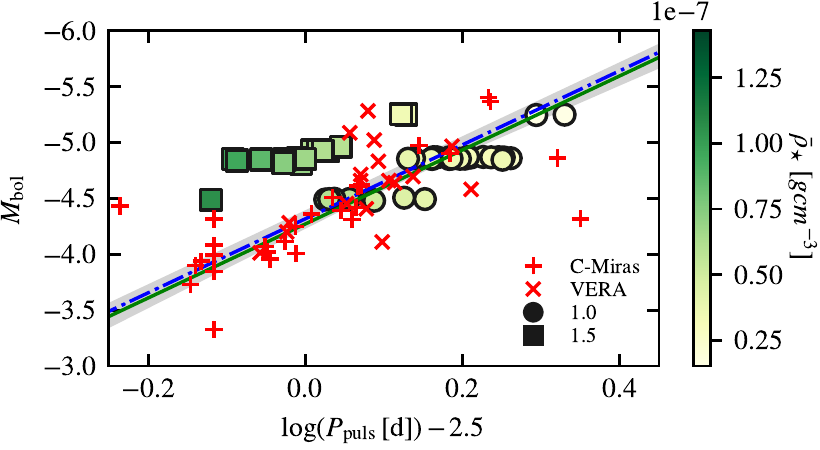}
      \caption{Extracted fundamental pulsation periods against the bolometric luminosity. The solid line is the P-L relation derived from the observed Carbon-rich Miras (plus-signs) in the LMC \citet{whitelockAsymptoticGiantBranch2009}; whereas the dash-dotted line is the P-L relation derived from Oxygen-rich Miras (x-marks) in the Milky Way \citep{andriantsaralazaplrelation2022}. The filled in grey region is the uncertainties in the empirical P-L relations (which overlap one another). Other symbols in the legend indicate the model stellar mass in $M_{\odot}$ of our AGB models.}
         \label{figplrel}
   \end{figure}

The P-L relation of AGB stars is an area that has been extensively studied and has been constrained by observations. Two independent P-L relations are compared with the 3D models for AGB stars, shown in Fig.~\ref{figplrel}. The first P-L relation was derived from observed carbon-rich Mira variables in the Large Magellanic Cloud (LMC), where a bolometric correction was applied in the K-band to derive the absolute bolometric luminosity \citep{whitelockAsymptoticGiantBranch2009}. \citet{andriantsaralazaplrelation2022} used improved constraints on the distances to oxygen-rich Mira variables in the Milky Way by making use of observations from Very Long Baseline Interferometry and masers of the AGB stars. They applied the radiative code \texttt{DUSTY} \citep{ivezicDUSTYRadiationTransport1999} to derive the bolometric luminosity of the sources. In Fig.~\ref{figplrel}, the 3D models are in relatively good agreement with both P-L relations, and within the parameter spread and scatter of the observed carbon and oxygen-rich Miras used to derive the respective P-L relations. There is evidence of the universality (that is one can use results from LMC sources in a different galactic environment) of a P-L relation for Miras \citep[see][]{alvarezOxygenrichMiraVariables1997, whitelockAGBVariablesMira2008, andriantsaralazaplrelation2022}. Nevertheless, one should be cautious in considering such universality, with counter arguments against it \citep[e.g.][]{barthesPeriodLuminosityColourDistributionClassification1999, uragoTrigonometricParallaxOrich2020}. The existence of some universality is suggested in Fig.~\ref{figplrel}. However, compared to the observations and their uncertainties, it is sensible to just conclude that our 3D models of AGB stars are within a realistic scatter of real AGB stars. 

%

The RSGs follow a different P-L relation, at the very least having different zero points in the K-band, though the slope may be in agreement with Miras \citep[e.g.][]{feastPeriodluminosityRelationSupergiant1980, woodLongperiodVariablesMagellanic1983, jurcevicPeriodluminosityRelationsRed2000, piercePeriodluminosityRelationsRed2000, kissVariabilityRedSupergiant2006, yangPERIODLUMINOSITYRELATIONRED2012, soraisamVariabilityRedSupergiants2018}. Existence of super AGB stars with $M_\textrm{bol}=-8$ \citep{poelarendsSupernovaChannelSuperAGB2008} also makes the limit between AGB stars and RSGs non-trivial. If the RSG models were to be included in Fig.~\ref{figplrel}, they would mostly populate the same $x$-axis range, but with $-6\ge M_\textrm{bol} \ge -8$. Previous studies suggest the existence of a universal P-L relation for RSGs; however, the uncertainties involved still allow for a wide range of parameter values for the slope and intercept in this P-L relation \citep[see Fig. 7 in][]{soraisamVariabilityRedSupergiants2018}. It is also risky to convert observations in the K-band into bolometric values for RSGs, as the colour correction formulations are more complicated and uncertain as the spectral evolution of the RSGs needs to be accounted for \citep[see][]{beasorEvolutionRedSupergiant2018, soraisamVariabilityRedSupergiants2018} and the distance estimates need to be good to obtain accurate corrections for extinction. Therefore, the comparison between the bolometric P-L relation for RSGs for our models against observations was not attempted here.
 
\subsection{Interaction between convection and pulsation} \label{sect_pulsationconvectioninteraction}
\subsubsection{Interpretation of the spread in the pulsation period}
 
\begin{figure}[t]
\centering
\includegraphics[width=\hsize]{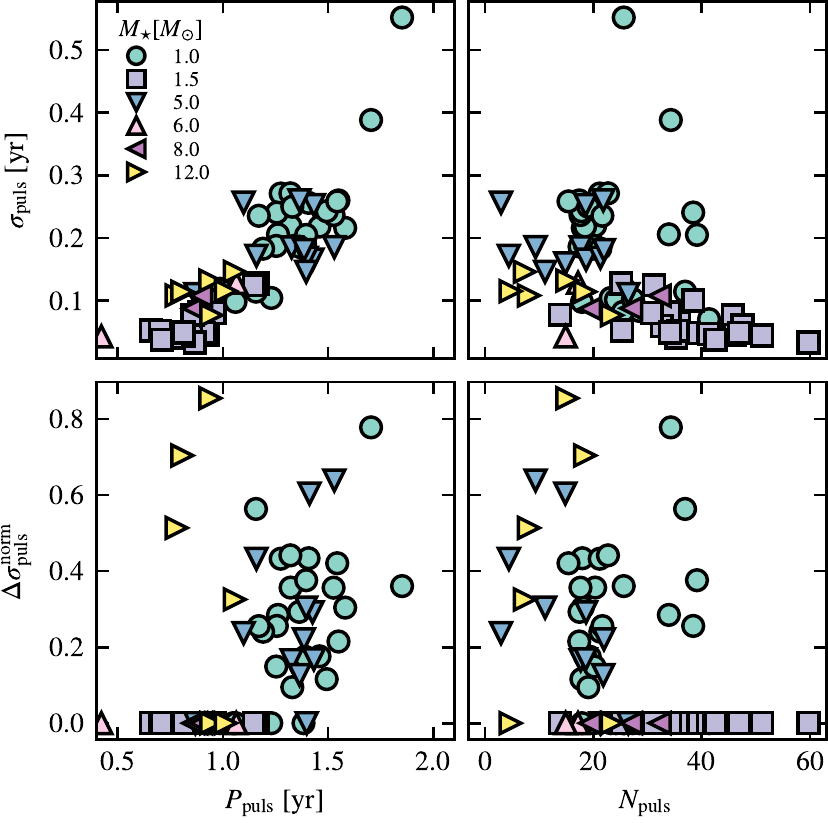}
  \caption{Dependence of the spread in the pulsation period on observed pulsation properties. The panels show the derived spread in the pulsation period (top row) and relative difference in the spread between using the relative or adaptive method (bottom row), against the pulsation period and the number of pulsations observed in the simulations (left and right columns, respectively).
          }
     \label{fig_periodunc_vs_sparams}
\end{figure}
%

Fluctuations in observed periods and luminosity at a given phase of a pulsation in cool red giants have been detected; several studies suggest that convection plays an important role in the fluctuations, or that the mechanism behind the fluctuations is stochastic in nature \citep[][]{eddingtonIrregularitiesPeriodLongperiod1929,percyLongTermChangesMira1999,kissVariabilityRedSupergiant2006,percyAmplitudeVariationsPulsating2013,cunhaSolarlikeMiraStars2020}. Furthermore, fluctuations can be expected when pulsations interact with the giant convective cells \citep[][]{schwarzschildScalePhotosphericConvection1975, antiaConvectionEnvelopesRed1984}. If this interaction manifests itself in the spread of the pulsation period,  $\sigma_\textrm{puls}$, the values of $\sigma_\textrm{puls}$ derived from the models (see Sect. \ref{sect_methodpulsation}) help to parameterise the fluctuations. Presented in Fig.~\ref{fig_periodunc_vs_sparams}, there is a good correlation between $P_{\textrm{puls}}$ and $\sigma_{\textrm{puls}}$. This correlation is to be expected since a longer $P_{\textrm{puls}}$ is associated with stars of lower surface gravity. \citet{freytagGlobal3DRadiationhydrodynamics2017} conclude the size of surface convection cells to be increasing with decreasing surface gravity in the AGB models. Therefore, larger surface convection cells and a more extended atmosphere, both consequently due to a lower surface gravity, could directly cause a larger spread in the pulsation period.

As a result of the positive correlation between $P_{\textrm{puls}}$ and $\sigma_{\textrm{puls}}$, the relations between $\sigma_{\textrm{puls}}$ and the fundamental stellar parameters follow closely with how $P_{\textrm{puls}}$ depends on the stellar parameters as in Fig.~\ref{fig_period_vs_sparams}. However, one important consideration is how $\sigma_{\textrm{puls}}$ would depend on the number of pulsations observed in the simulation, $N_\textrm{puls}$. For the period determination process to be reliable, a sufficiently high number of pulsations need to be considered so that the $\sigma_{\textrm{puls}}$ value converges to a final value. As seen in Fig.~\ref{Figcomparetypes}, the $1M_{\odot}$ model had multiple strong peaks in its FFT spectrum, resulting in a larger $\sigma_{\textrm{puls}}$. Values of $\sigma_{\textrm{puls}}$ do not necessarily decrease with more observed pulsation cycles in the simulations. Primarily this indicates the fluctuations are consistent throughout the simulations. The relative difference in the $\sigma_\textrm{puls}$, whether using a relative or adaptive width to do the weighted averaging in the period determination process, was calculated with $\Delta \sigma_\textrm{puls}^\textrm{norm}=(\sigma_\textrm{puls}^\textrm{Ada}-\sigma_\textrm{puls}^\textrm{Rel})/\sigma_\textrm{puls}^\textrm{Rel}$ and was also plotted against $P_{\textrm{puls}}$ and $N_{\textrm{puls}}$. Where $\Delta \sigma_\textrm{puls}$ is large in the bottom row of Fig.~\ref{fig_periodunc_vs_sparams}, it provides further indication of the interplay between convection and pulsations, as the adaptive width method captures signals surrounding the pulsation signal which are missed by the relative width method. For the $1\,M_{\odot}$ AGB stars, convection can lead to substantial signals being present in the FFT power spectrum, in part due to the AGB stars having low surface gravity. On the other hand, RSGs with high $\Delta \sigma_\textrm{puls}$ are indicative of lower pulsation amplitude. Signals due the interference of convection on the pulsations become relatively more appreciable hence would lead to higher $\sigma_\textrm{puls}$ using the adaptive width method. In this way, Fig.~\ref{fig_periodunc_vs_sparams} captures an important point that in our models, it is clear that convection and pulsations do interact continuously.

\subsubsection{Regularity or irregularity of the pulsations}

\begin{figure}[t]
\centering
\includegraphics[width=\hsize]{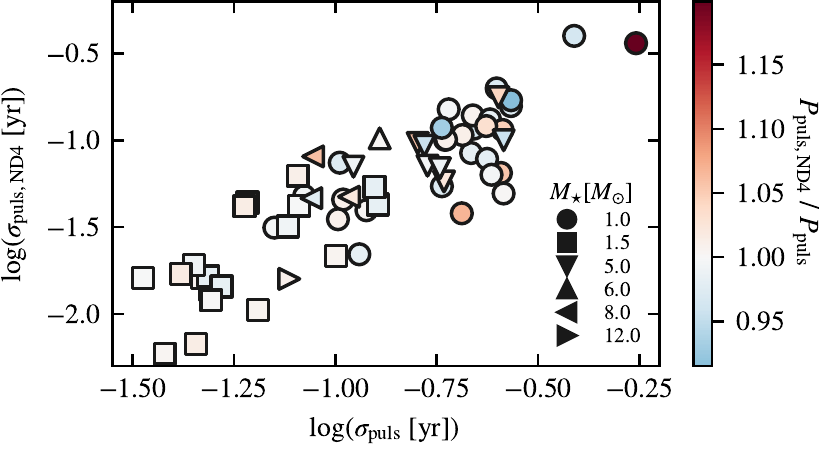}
    \caption{Standard deviation in the four periods calculated when the sequence was split into quarters against the spread of the period in the whole sequence. The ratio of the averaged period across the split quarters and the full sequence is coloured in. A similar plot was obtained when the full sequence was split into halves.}
     \label{figsplitter}
\end{figure}

For models with low values of $\sigma_{\textrm{puls}}$ in Fig.~\ref{fig_periodunc_vs_sparams} it does not mean that there is no variability in the pulsations being observed. Examining instances of the regions where the pulsation amplitude significantly decreases in Fig.~\ref{Figcomparetypes}, what follows such events can be a slight phase shift in the pulsations and variation in the amplitude. Thus this may contribute to non-zero values of $\sigma_{\textrm{puls}}$. Another way to identify the variability explicitly was to slice the simulation of models into parts, and repeat the pulsation period extraction procedure on the split parts. This was done on models with at least $3000$ time steps. The resulting derived periods are represented in Fig.~\ref{figsplitter}, where a good agreement exists between the scatter of the period in the split sequences against the spread of the period in the full sequence.

It is notable that apart from a few extreme cases, the averaged pulsation period across the split sequences agree with the period derived from the full sequence to about $5\,\%$. In addition, it was verified that almost all of the pulsation periods from the split sequences are within the main spread in the period of the full sequence, and when outside, their respective period spreads overlap with each other. The lack of outliers primarily indicates that the pulsation extraction process is reliable, considering the degree of irregularity in the dynamics of the pulsating star. It also shows good regularity of the pulsations despite the variability. Where there are large values in the spread of the period in any of the split sequences, it suggests an event that affects the pulsations. An example of this is the sudden decreases in the pulsation amplitudes observed in the representative models seen in Fig.~\ref{Figcomparetypes}. However, this again highlights the interaction between convection and pulsations, where in certain intervals of the simulation, fluctuations and changes in the period were detected. 

%

\subsection{Stellar radius parameter} \label{sect_radiustest}

\subsubsection{Entropy minimum as the radius definition} \label{sect_radiuscomparison}

In the analyses presented in this work, choosing a consistent definition for the radius was extremely important, as several physical parameters and properties were derived from it. The multiple chosen definitions of the radius were introduced in Sect. \ref{sect_methodradiusdef}. Barring the radius estimated from the phase of the pulsation, a sound-crossing timescale was computed for each radius definition. This sound-crossing time was defined as the time it takes for a sound wave to travel back and forth from the centre to the defined radius of the star. We assume the sound-crossing time to be a good first-order estimation of the pulsation period, and first verified that the two timescales correlate linearly to each other. We can equivalently expect the free-fall timescale (over a substantial part of the radius of the star) to be in reasonable agreement with both the sound-crossing and pulsation timescales; by considering the virial theorem \citep{christensen-dalsgaardPulsationTheoryStellar1993}.

\begin{figure}[t]
\centering
\includegraphics[width=\hsize]{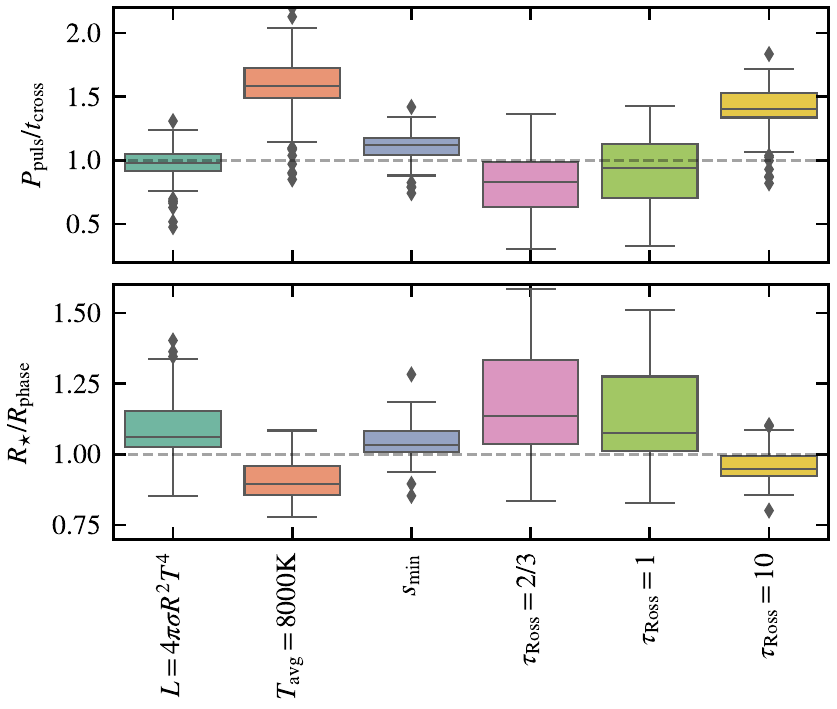}
  \caption{Box plots investigating the choice of stellar radius. \textit{Top panel}: Box plot of the ratio of the pulsation period and sound-crossing time. \textit{Bottom panel}: Box plot of the ratio of the stellar radius and the radius below which the layers are largely in phase. The sound-crossing times and stellar radii are from the six radii definitions tested in the models presented in this work as outlined in Sect. \ref{sect_methodradiusdef}. Each box plot represents the distribution by drawing the box from first to third quartiles ($Q_1$ and $Q_3$, respectively) with a horizontal line drawn in between to denote the median, and the height of the box is the interquartile range (IQR). The bottom and top whiskers are bounded by bars denoting the minimum and maximum values, defined as the lowest and highest values within the range $\textrm{min} \ge 1.5\textrm{IQR}-Q_1$ and $\textrm{max} \le 1.5\textrm{IQR}+Q_3$, respectively. Outliers are identified as values being outside of the whisker bounds, denoted by the rhombus symbols in the plot. 
          }
     \label{figradfits}
\end{figure}

The discrepancies in the pulsation periods derived from using the different radius definitions (since the location of the thin layer being considered is different in the weighted averaging, see Sect. \ref{sect_mainmethod}) were verified to be minimal. The ratio between the pulsation period and the computed sound-crossing times from the different definitions are presented in the top panel of Fig.~\ref{figradfits} in the form of box plots.  While we expect the sound-crossing times to be a good approximation for the pulsation periods, we note the median values of the box plots in the top panel of Fig.~\ref{figradfits} are surprisingly close to $1$ -- and not with a systematic offset from $1$ as initially expected. Ideally, the box plots should have a thin interquartile range with minimal outliers. The entropy minimum definition best matches this case. Using the average temperature of $8000\,$K and Rosseland optical depths of $10$ underestimate the extent of the radius, whereas using Rosseland optical depths of $2/3$ and $1$ overestimate the radius and the respective box plots have relatively large interquartile ranges. The radius definition used in \citet{freytagGlobal3DRadiationhydrodynamics2017} fared well in this test; however, there were outliers which can be attributed to the spatial fluctuations in density and opacity in the stellar atmosphere, which can be mistaken as the boundary for the stellar radius. %
%

To further investigate these statements, box plots of the ratio between the radius derived from the different radius definitions and the averaged radius where the pulsations are in phase are presented in the bottom panel of Fig.~\ref{figradfits}. We find qualitative agreement and draw the same conclusions with the box plots in the top panel of Fig.~\ref{figradfits}. Additionally, linear regressions were applied on the relationships being analysed in Fig.~\ref{figradfits}; the ideal case for the linear regression would simply be (for $Y=a+bX$) $a=0$ and $b=1$, with $R^{2}=1$. The entropy minimum definition best matched this case in both relationships. These results reaffirm that the radius, as defined by the entropy minimum, was the most consistent and reliable method to calculate the radius in the models. Currently there is no equivalent definition available from an observational stand point. In terms of pulsations, however, we were interested in capturing the outer boundary containing most of the dynamical action. The radius as determined by the entropy minimum definition appears to be the best option in our models. %

\subsubsection{Radius amplitude due to pulsations}

\begin{figure}[t]
\centering
\includegraphics[width=\hsize]{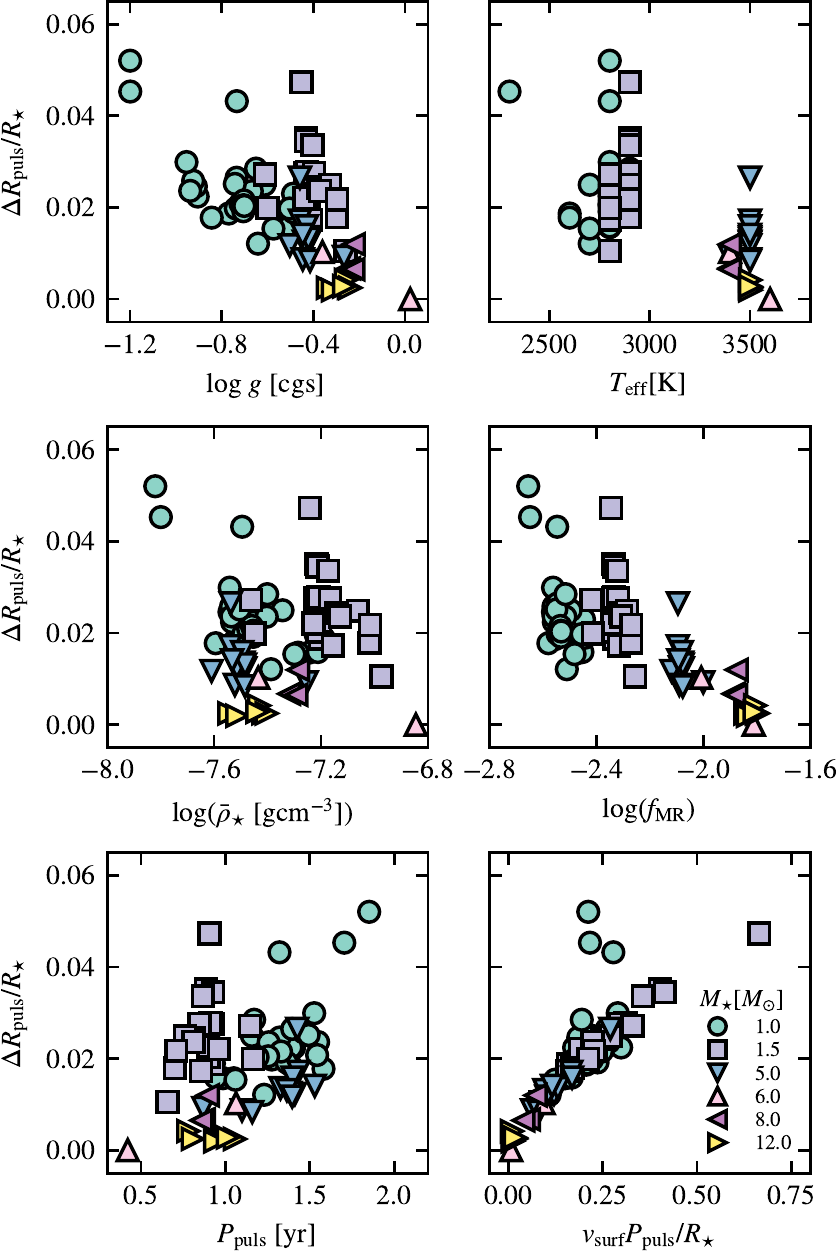}
  \caption{Effect of the pulsations on the stellar radius. The panels show the relative amplitude due to pulsations against (from top to bottom reading left to right): surface gravity; effective temperature; mean density; mass-to-radius ratio; pulsation period and the surface velocity multiplied by the pulsation period over the stellar radius.
          }
     \label{figradamp}
\end{figure}   
%

The standard deviation of the radius variation in the entropy minimum definition of the radius, displayed in the right panel of Fig.~\ref{entropydef}, was calculated and taken as the pulsation amplitude, $\Delta R_\textrm{puls}$. The corresponding ratio $\Delta R_\textrm{puls} / R_{\star}$, hereby called the relative amplitude, is plotted against various parameters in Fig.~\ref{figradamp}. In general, the relative amplitude follows the trends set out by previous analyses and seems to react strongly to the surface gravity. Against $P_\textrm{puls}$, the relative amplitude increases with longer pulsation periods (associated with low surface gravity). Following this, the relative amplitude increases with decreasing values of surface gravity, mean-density, and mass-to-radius ratio, specifically for each mass class of the models. It is also worth pointing out that despite low relative amplitudes, the RSGs still produce shocks and strong outward flows in the outer atmosphere.

The relative amplitude increases almost linearly with the quantity $v_\textrm{surf}P_\textrm{puls}/R_\star$, where $v_\textrm{surf}$ is the surface velocity, calculated as the standard deviation value (over time) of the spherically averaged velocity at the radius. This quantifies how well the model represents a perfectly spherically symmetric model with strong radial pulsations. In the bottom right panel of Fig.~\ref{figradamp}, the three AGB stars that deviate from the linear relation are of low surface gravity with very extended atmospheres. The pulsations, together with the low surface gravity and extended atmosphere, enhance the likelihood for the atmosphere to deviate further away from spherical symmetry.

\section{Discussion} \label{sect_maindiscussions}

\subsection{Pulsation constant parameter}

Values of $Q$ may help to characterise different pulsation modes (for example, radial or non-radial, or, fundamental or overtone modes) existing in a star. Given its dependence on the stellar parameters shown in Eq.~(\ref{equation_pulsationconstant}) and for a given period, different values of $Q$ indirectly differentiate the stellar interior, especially with regards to the temperature stratification in the star. As suggested in \citet{kissVariabilityRedSupergiant2006} however it could be useful to use $Q$ to verify or compare stellar parameters as a check between theoretical models and observations.

Works to derive theoretical values of $Q$ have been produced through usage of 1D models. \citet{xiongNonadiabaticOscillationsRed2007} derived the values of $Q$ analytically from pulsations prescribed by non-local time-dependent convection in 1D for AGB stars ($M_{\star} \le 3M_{\odot}$). From their models, the $Q$ values ranged between $0.06$ and $0.1$ days, with an average of $0.08$ and a $1{\text -}\sigma$ scatter of $0.077 \pm 0.01$ days. In comparison, the $1{\text -}\sigma$ scatters for our models of AGB stars, RSGs and as a whole were respectively calculated to be: $Q^\textrm{AGB}=0.0724\pm0.0049$, $Q^\textrm{RSG}=0.0644\pm0.0105$ and $Q^\textrm{all}=0.0699\pm0.008$ days. \citet{xiongNonadiabaticOscillationsRed2007} had a considerable amount of scatter in the values of $Q$ \citep[see Fig. 1 of][]{takeutiMethodEstimateMasses2013}, but there was a tight correlation between the $Q$ values in the fundamental and overtone modes of their models. \citet{takeutiMethodEstimateMasses2013} makes use of the tight correlation to constrain masses of AGB stars with observed radii, thus presenting a new method to derive masses of observed AGB stars. For this method to be reliable, however, it requires a relatively certain measure of the observed radii and that the theoretical AGB models pulsate in more than one mode (in the fundamental and at least one overtone mode). The lack of overtones in our AGB models means it is not currently possible to apply a similar method and compare with the estimates for the masses of the AGB stars in \citet{takeutiMethodEstimateMasses2013}.

\begin{figure}[t]
   \centering
   \includegraphics[width=\hsize]{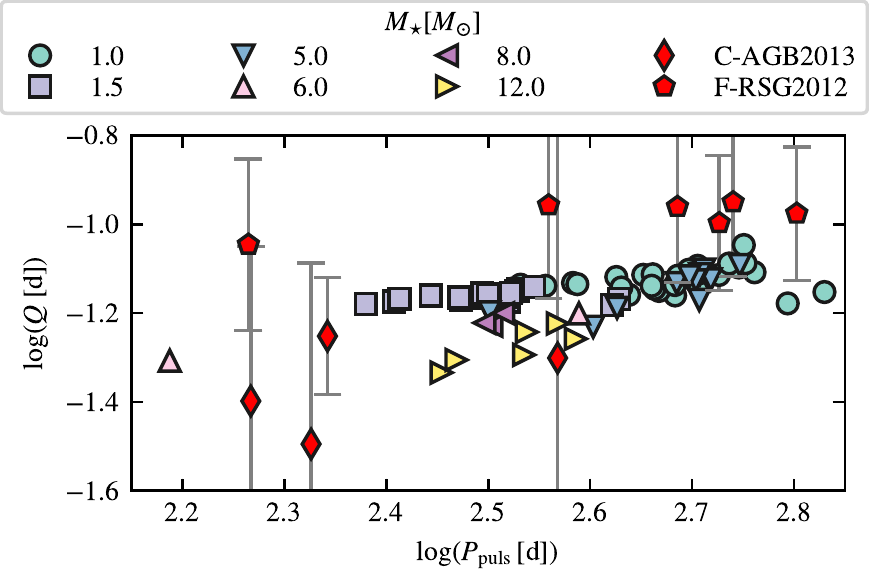}
      \caption{Logarithm of the pulsation constant, $Q$,  against the logarithm of pulsation period. To compare against the models, $4$ Mira variables (rhombuses) and $7$ RSGs (pentagons) are included from \citet{cruzalebesFundamentalParameters162013} and \citet{fadeyevNonlinearPulsationsRed2012}, respectively. Uncertainty bars have been excluded for the models to preserve the clarity of the plots.
              }
         \label{figpulsationconst}
   \end{figure}

\citet{cruzalebesFundamentalParameters162013} derive values of $Q$ for a number of late-type giants. They used observations from the VLTI/AMBER facility (in the K-band) and the \texttt{MARCS} stellar atmosphere models \citep{gustafssonGridMARCSModel2008} to derive new values of angular diameters with higher certainty than in previous studies. The initial input stellar parameters for the \texttt{MARCS} models were effective temperature, surface gravity and stellar mass adopted from the spectral type of the stars in their sample. Due to the multiple options of using different spectral types to infer the input stellar parameters, they confirmed that the derived angular diameters were not significantly affected by the choice of input stellar parameters. Finally, the stellar mass of the targets in their sample was estimated by correlating the temperature and luminosity acquired from the \texttt{MARCS} models to evolutionary tracks in the H-R diagram. From their sample, there were only 4 Mira variables in the fundamental pulsation mode which could be compared to the AGB stars in the present work, which are over-plotted in Fig.~\ref{figpulsationconst}. 

For RSGs, \citet{fadeyevNonlinearPulsationsRed2012} uses the solution of radiation-hydrodynamics and turbulent time-dependent convection equations to study non-linear oscillations in RSGs with the zero age main sequence (ZAMS) masses in the range of $8M_{\odot} \le M_{\textrm{ZAMS}} \le 20_{\odot}$. From their set of models, the relation of their corresponding plot of $Q$ to the stellar mass-to-radius ratio provides the relation $\log Q=-2.288-0.778\log f_{\textrm{MR}}$. We derive the same relation in from our RSG models in this work ($M_{\star} \ge 5M_{\odot}$) being $\log Q=-(2.238 \pm 0.177)-(0.527 \pm 0.094)\log f_{\textrm{MR}}$, where the $\log Q-\log f_{\textrm{MR}}$ relation is reflected in the bottom row of Fig.~\ref{figrhomr}.   \citet{fadeyevNonlinearPulsationsRed2012} further extended the use of their derived $Q$ to estimate the stellar masses of $7$ galactic RSGs with available periods. They made use of the radii determined from \texttt{MARCS} models in \citet{levesqueEffectiveTemperatureScale2005} and \citet{josselinAtmosphericDynamicsMass2007} and acquired the corresponding stellar masses using their values of $Q$. From their derived stellar parameters, the corresponding $Q$ values (for RSGs with pulsation periods higher than 100 days) were calculated in order to be compared to our models, and are over-plotted accordingly in Fig.~\ref{figpulsationconst}. 

%

While there is some overlap, Fig.~\ref{figpulsationconst} overall shows there is  discrepancy between the estimates of $Q$ of observed stars provided by \citet{cruzalebesFundamentalParameters162013} and \citet{fadeyevNonlinearPulsationsRed2012}, and the models in this present work. Different theoretical models being compared, alongside observational uncertainties involved, contribute to the discrepancies. Furthermore, factoring in the high uncertainties in the radius estimations inferred from observations, as well as further differences in the definition of the radius of a star (see Sects. \ref{sect_radiuscomparison} and \ref{sect_radiusdiscuss}), the discrepancies between our work here and the values derived from observations in \citet{cruzalebesFundamentalParameters162013} and \citet{fadeyevNonlinearPulsationsRed2012} are to an extent unavoidable. Both of the works used the Rosseland radius definition ($\tau_\textrm{Ross}=1$) in the \texttt{MARCS} models to define the radii being used. However, any radii calculated via the \texttt{MARCS} models should be taken with a level of caution, as \texttt{MARCS} assumes static atmospheres, whereas the stars being compared here are dynamical in nature, evident from the pulsations. Hence it is challenging to compare the works here, especially as the parameters were indeed derived from different methods, which involves their own uncertainties; however, Fig.~\ref{figpulsationconst} represents an exploratory attempt to find comparisons in literature and observations, and show that the pulsation constants of our models are within the general range derived from available observations.

\subsection{Difference in models} \label{sect_discussionmodeldifferences}
Section \ref{sect_modelsstetup} introduced the general setup of the models, further developed from previous works in 3D simulations of pulsating AGB stars \citep[see][]{freytagThreedimensionalSimulationsAtmosphere2008, freytagGlobal3DRadiationhydrodynamics2017}. As part of the work presented here, pulsations extracted from different generations of the \texttt{CO5BOLD} code were investigated. Higher resolution models introduced additional finer convective structures \citep{freytagGlobal3DRadiationhydrodynamics2017} and slightly more efficient convection in the models, but the period of the pulsations was consistent with lower resolution models. Temporally, longer running simulations help to investigate the fluctuations in the pulsations more reliably -- and smaller sampling time steps do not reveal anything new.

The inclusion of radiative pressure in the stellar interior becomes more important with higher stellar mass and inner temperatures \citep[see Sect. \ref{sect_modelsstetup} and][]{goldbergNumericalSimulationsConvective2022}. After examining the radiation-to-gas pressure ratio in the interior, the missing radiative pressure is expected to affect the deep stellar interior of RSGs, and could enhance convectively unstable regions in the core of the RSG models. In Figs. \ref{figrhomr} and \ref{figpulsationconst}, there is a $6 M_{\odot}$ RSG model that seemingly identifies as an outlier when compared to other RSGs. This model (\texttt{st36g00n04}) was one of the first attempts to produce a representative model of Alpha Orionis (Betelgeuse). The model has realistic temperature and surface gravity but smaller mass and radius than the inferred stellar parameters of Betelgeuse. Temperature and gravity are at the upper end of our model set (see Fig.~\ref{fig_period_vs_sparams}.), while the numerical resolution is comparatively low.
This leads to relatively inefficient convection not carrying energy up from the core region, where radiative energy transport dominates, but only in the outer layers of the star. The entropy profile formally indicates convective stability of the inner region, while overshooting convective velocities still can reach very deep down. In this case, we believe that the inclusion of radiation pressure would deepen the convection zone.
The relatively shallow convection zone might have consequences for the excitation of the pulsations. But it should have only a minor impact on the pulsation period, because for the period the outer layers, where the sound speed is low,
are of dominating influence.
Thus, \texttt{st36g00n04} serves as an extreme case to highlight possible limitations in our current set of RSG models. However, it was verified that this model, too, is pulsating in the fundamental mode. Hence, its results still appear valid and useful to maximise the range in our parameter space in the analyses presented in this work.

The current generation of models have only explored a relatively tight region of parameter space in the mass-to-radius ratio, $f_{\textrm{MR}}$ (see Sect. \ref{rittersmeandensity}), in each mass class, indicated in the bottom panel of Fig.~\ref{figrhomr}. The reasoning for this is different for the AGB and RSG models. For AGB models, the choice of stellar parameters was partially guided by DARWIN models that produce stellar winds driven by radiation pressure on dust \citep[see][]{erikssonSyntheticPhotometryCarbonrich2014, bladhCarbonStarWind2019, bladhExtensiveGridDARWIN2019}. As the escape velocity from a star scales with  $v_{\textrm{esc}}\propto(M_\star/R_\star)^{0.5}$, a lower $f_{\textrm{MR}}$ was preferred to generate stellar winds easily. This is since matter can be lifted up to distances from the star, where low temperatures permit dust condensation. In global RSG models, consideration towards the size of the convection cells at the surface and the ability to resolve them becomes important; where the size of granules, or convection cells, is proportional to the pressure scale height \citep[see for instance][and references therein]{freytagScalePhotosphericConvection1997}. Larger pressure scale heights are then preferred since this would allow surface features to be resolvable. In relation to the ratio of the radius and pressure scale height at the photosphere: $R_{\star}/H$, with $H \propto 1/g \propto R_{\star}^{2}/M_{\star}$,  the ratio becomes $R_{\star}/H \propto M_{\star}/R_{\star}$. Therefore, the preference of larger pressure scale heights has limited the ranges of the $f_{\textrm{MR}}$ in simulations of the RSGs. In general however, going towards higher ratios of $f_{\textrm{MR}}$ requires higher spatial resolution, which is more computationally expensive. 

\subsection{Convection and pulsation interaction}

Section \ref{sect_pulsationconvectioninteraction} highlighted the evidence of the interaction between convection and pulsations in the models. This interaction has several effects on the atmosphere, importantly creating inhomogeneities on the surface of the photospheric layer. 

The surface inhomogeneities due to convection may interact with the pulsations and atmospheric shock waves. This interaction produces a network of shock waves in the upper atmosphere \citep{freytagGlobal3DRadiationhydrodynamics2017}. In this sense, the pulsations and shock waves can carry forwards any spatial inhomogeneities on the surface, creating patchy and irregular outflows. Thus it is thought the size of surface convective cells and how the pulsations interact with convection are significant factors for the production of irregular and patchy outflows. The interaction also causes variability in the pulsation period and amplitude.

While the study of the interaction was not presented here, it is evident decoupling the pulsations and convection is difficult. The reason for the difficulty in investigating the non-linear interplay between convection and pulsations is that there is no explicit parameterisation for how the convection affects the pulsation, and vice versa, in the \texttt{CO5BOLD} code. Convective signals still remain substantial despite applying spherical averages to obtain the radial velocity: observed in the first and third rows of Fig.~\ref{Figcomparetypes}. In fact, the pulsations heavily modulate the convection energy flux (the sum of the enthalpy and kinetic energy fluxes) at every pulsation cycle. Reacting against this, the stratification in the stellar interior and atmosphere may experience perturbations in the form of pressure fluctuations. The mass, momentum and energy fluxes are affected, and the degree of this disturbance can influence the global behaviour of the pulsations. We interpret sudden amplitude change in the pulsations, as observed for the $1.5 \, M_{\odot}$ and $8 \, M_{\odot}$ models in the top row Fig.~\ref{Figmainmethod}, to be due to interference by convection. Given the high Mach numbers (velocities exceeding $1\,\textrm{kms}^{-1}$) of the convective flows and modulation due to pulsations, quantifying the interaction between convection and pulsations remains challenging. 

Figures \ref{fig_periodunc_vs_sparams} and \ref{figsplitter} investigate the relationship between fluctuations in the pulsation period and stellar parameters. The investigations represent a phenomenological way to study the effect of convection on the pulsation period. For models that experience regular, high amplitude pulsations (interpreted as models with low $\sigma_\textrm{puls}$), convection still causes some variability in the period following events where pulsation amplitudes suddenly decrease. Capturing the variations and asymmetries due to the non-linear interaction between convection and pulsations is important in reproducing more realistic dynamical models and the mass loss mechanism \citep[see][]{liljegrenDustdrivenWindsAGB2016}.

\subsection{Stellar radius definition} \label{sect_radiusdiscuss}

Works determining the radius of AGB stars and RSGs from observations fundamentally rely on how the radius, that is the boundary of the star, is first defined. Historically there have been competing ideas on which definitions work best, especially for stars with cool extended envelopes \citep{haniffNewDiameterMeasurements1995}. Furthermore, the extent of the surface of cool red giants carries high uncertainty due to time-variability, non-sphericity, and extensions of the surface of the star arising from the dynamical atmosphere \citep{arroyo-torresWhatCausesLarge2015}. Therefore, it becomes non-trivial to decide how to define the radius from observations and how to compare with the radii obtained in theoretical models. 

It is difficult to compare the radii of AGB stars and RSGs obtained from the \texttt{CO5BOLD} simulations against radii estimated from observations. Fundamentally, this comes down to the radius definition being used as suggested in Sect. \ref{sect_radiuscomparison}. From the observational view, the dynamical nature of the atmosphere and the molecular and dust formation shells surrounding cool evolved stars \citep[e.g.][]{wittkowskiMiraVariableOrionis2007, khouriStudyInnerDust2016, ohnakaClumpyDustClouds2017, hofnerMassLossStars2018} evidently present themselves as uncertainties to the true radius of the star. Analogous to the different conditions which currently exist to describe the radius of the star in the \texttt{CO5BOLD} models, radii from the observational stand point also have various definitions, being mostly dependent on the wavelength observed and what is possible technologically. The distance measurement to target sources is also a major uncertainty in the observed radii of cool red giants \citep[see][]{baschekParametersTeffStellar1991, haniffNewDiameterMeasurements1995, chiavassaOpticalInterferometryGaia2020}. Despite the complications, it is however reassuring that for the AGB models, Fig.~\ref{figplrel} shows that the over-plotted P-L relations are in good agreement with the scatter of the observed Miras, indicating that the stellar parameters in, or acquired from, the models are in the region of those of real stars. 

Section \ref{sect_radiustest} investigated which definition for the radius yielded the most reliable radius of the modelled star. It is important to state that our primary concern was to decide on a definition that would capture the shell where most of the dynamic action (contained within the star), particularly pulsations, took place. The entropy minimum definition, marking the region where the atmosphere becomes convectively stable, appeared to be the best definition in Fig \ref{figradfits}. Several different stellar parameters, such as the effective temperature, mean-density and mass-to-radius ratio, can be derived from the radius definitions, hence finding the most reliable definition is crucial. Using the entropy minimum definition also yielded the closest agreement to the $P_\textrm{puls}-\bar{\rho}_{\star}^{-0.5}$ relation as in Fig.~\ref{figrhomr}. This reassures that the radius definition we adopted is the best option amongst the other definitions presented in this work. The closest equivalent observational definition of the radius to this would be the very inner boundary of the photometric radius of the observed star. This requires either a combination of high precision interferometry and distance estimates to the star itself, or precise radiative transfer modelling of the luminosity and temperature of the star to obtain estimates for the photometric radius of the star. 

\section{Summary and conclusions} \label{sect_mainconclusions}
Global 3D radiation-hydrodynamics models of AGB and RSG stars are an essential tool for studying their pulsation properties and the excitation mechanism. The 3D models computed with the \texttt{CO5BOLD} code were used to calculate the fundamental pulsation frequency and other critical pulsation properties. A total of $75$ models was analysed in this work, comprising several generations of models. The convection and pulsations are both emergent properties of the radiation-hydrodynamics simulations. This means that no parametric treatment of convective energy transport is necessary, in contrast to classical 1D pulsation models. Despite the non-linear nature of the interplay between convection and pulsations, in all of the models radial pulsations can clearly be observed and the pulsation periods in the fundamental mode were extracted. A good correlation was found for the $P_\textrm{puls}-\bar{\rho}_{\star}^{-0.5}$ relation in the models, confirming that the pulsations are in the fundamental radial mode. This agrees with current theory that for LPVs, the variability is attributed to low-order large-amplitude pulsations, existing as standing waves. 

In their dependence on fundamental stellar parameters, the pulsation periods agree with expectation, such as a lower surface gravity resulting in longer periods. The AGB models agree well with available observations in the P-L relation, indicating that the stellar parameters and the pulsation properties are realistic. 

Applying regression models led to the mass estimate formula in Eq.~(\ref{equation:mass_estimate}), as a function of the fundamental pulsation period and stellar radius. The radius value to be used in Eq.~(\ref{equation:mass_estimate}) must be consistent with the radius used when the formula was derived. We found that identifying the innermost local entropy minimum in the atmosphere provided the best measure for the stellar radius in the context of pulsation analysis. The purpose of identifying the best radius definition was to explicitly capture the boundary of dynamics in the stellar interior, thus where the pulsations play a dominant role.

It is clear from the simulations that convection actively affects the pulsation and that their interplay causes period variability. There is evidence that a reverse effect exists, in the sense that the pulsations also impact the convection. The deep convective regions and large convection cells characteristic of evolved red giants make it important to understand the interplay with pulsation, and ultimately how the interplay affects the massive stellar winds and the mass loss process. Further and more in-depth studies on the interaction of convection and pulsations in the 3D models remain on the agenda for future work. It proves to be a difficult topic to investigate, since the effect of one cannot be isolated, or decoupled, from the other.

\begin{acknowledgements}
      This work is part of a project that has received funding from the European Research Council (ERC) under the European Union’s Horizon 2020 research and innovation programme (Grant agreement No. 883867, project EXWINGS) and the Swedish Research Council (Vetenskapsrådet, grant number 2019-04059). Most of the computations were performed on resources provided by the Swedish National Infrastructure for Computing (SNIC) at UPPMAX and NSC in Sweden, supported by the Swedish Research Council (Vetenskapsrådet). Some simulations were performed on resources provided by PSMN in Lyon, France, and MPA in Garching, Germany.
\end{acknowledgements}

%
%

\bibliographystyle{aa_url}
\bibliography{aabfsh_pulsprop_accepted_arxiv01} 

\onecolumn 
\begin{appendix}
\section{Model parameters and results}%
    \begin{table*}[!htbp]
        \caption{Parameters and pulsation properties of models used in this work.}
        \label{table:appendixtablemodel}
        \centering
        \begin{tabular}{lcccccccccccc}
            \hline \hline
Model & $M_\star$ & $L_\star$ & $n_x^3$ & $x_\mathrm{box}$ & $n_t$ & $t_\mathrm{avg}$ & $R_{\star,s_\mathrm{min}}$ & $T_{\mathrm{eff},s_\mathrm{min}}$ & $\log g_{s_\mathrm{min}}$ & $P_\mathrm{puls}$ & $\sigma_\mathrm{puls}$ \\
 & $M_\sun$ & $L_\sun$ &  & $R_\sun$ &  & yr & $R_\sun$ & K & (cgs) & yr & yr \\
\hline
st26gm07n001 &  1.0 &  6953 & 281$^3$ & 1381 &  3502 & 27.74 &  358 & 2783 & -0.675 & 1.383 & 0.183 \\
st26gm07n002 &  1.0 &  6978 & 281$^3$ & 1381 &  3601 & 28.52 &  380 & 2706 & -0.725 & 1.581 & 0.216 \\
st27gm06n001 &  1.0 &  4982 & 281$^3$ & 1381 &  3602 & 28.53 &  324 & 2693 & -0.588 & 1.230 & 0.105 \\
st28gm05n001 &  1.0 &  4990 & 281$^3$ & 1381 &  3202 & 25.36 &  291 & 2846 & -0.494 & 0.986 & 0.119 \\
st28gm05n002 &  1.0 &  4978 & 281$^3$ & 1381 &  3201 & 25.35 &  300 & 2799 & -0.521 & 1.050 & 0.102 \\
st28gm05n003 &  1.0 &  4997 & 317$^3$ & 1263 &  3393 & 26.87 &  283 & 2882 & -0.472 & 0.943 & 0.084 \\
st28gm05n004 &  1.0 &  4961 & 317$^3$ & 1263 &  2993 & 23.70 &  279 & 2901 & -0.456 & 0.932 & 0.082 \\
st28gm05n005 &  1.0 &  4964 & 317$^3$ & 1263 &  4795 & 37.98 &  279 & 2897 & -0.459 & 0.918 & 0.071 \\
st28gm05n006 &  1.5 &  4956 & 317$^3$ & 1263 &  2099 & 16.62 &  270 & 2946 & -0.253 & 0.657 & 0.053 \\
st28gm05n008 &  1.0 &  4914 & 317$^3$ & 1263 &  2394 & 18.96 &  303 & 2775 & -0.529 & 1.060 & 0.098 \\
st28gm05n009 &  1.5 &  7452 & 317$^3$ & 1263 &  4103 & 32.50 &  332 & 2943 & -0.431 & 0.907 & 0.064 \\
st28gm05n011 &  1.5 &  7439 & 317$^3$ & 1263 &  3804 & 30.13 &  327 & 2966 & -0.417 & 0.885 & 0.081 \\
st28gm05n012 &  1.5 &  7424 & 317$^3$ & 1263 &  4604 & 36.47 &  326 & 2969 & -0.415 & 0.883 & 0.047 \\
st28gm05n014 &  1.5 &  7419 & 317$^3$ & 1263 &  4494 & 35.59 &  322 & 2983 & -0.406 & 0.926 & 0.100 \\
st28gm05n016 &  1.5 &  7474 & 317$^3$ & 1263 &  5194 & 41.14 &  325 & 2978 & -0.414 & 0.900 & 0.076 \\
st28gm05n017 &  1.5 &  7447 & 317$^3$ & 1263 &  5394 & 42.72 &  322 & 2986 & -0.408 & 0.899 & 0.061 \\
st28gm05n018 &  1.5 &  7472 & 317$^3$ & 1263 &  3804 & 30.13 &  324 & 2981 & -0.412 & 0.924 & 0.060 \\
st28gm05n019 &  1.5 &  7434 & 317$^3$ & 1263 &  5394 & 42.72 &  325 & 2974 & -0.414 & 0.927 & 0.045 \\
st28gm05n020 &  1.5 &  7664 & 317$^3$ & 1263 &  3734 & 29.57 &  328 & 2979 & -0.424 & 0.961 & 0.081 \\
st28gm05n021 &  1.5 &  7424 & 317$^3$ & 1263 &  5395 & 42.73 &  324 & 2977 & -0.411 & 0.912 & 0.049 \\
st28gm05n022 &  1.0 &  5039 & 317$^3$ & 1263 &  5395 & 42.73 &  314 & 2746 & -0.558 & 1.157 & 0.115 \\
st28gm05n023 &  1.5 &  6957 & 317$^3$ & 1263 &  3995 & 31.64 &  313 & 2978 & -0.383 & 0.848 & 0.049 \\
st28gm05n024 &  1.5 &  6646 & 397$^3$ & 1581 &  1491 & 11.80 &  311 & 2956 & -0.376 & 0.855 & 0.078 \\
st28gm05n025 &  1.5 &  6946 & 317$^3$ & 1263 &  6533 & 51.75 &  315 & 2970 & -0.387 & 0.866 & 0.034 \\
st28gm05n028 &  1.5 &  6907 & 317$^3$ & 1263 &  5264 & 41.69 &  308 & 2999 & -0.368 & 0.815 & 0.045 \\
st28gm05n029 &  1.5 &  6901 & 477$^3$ & 1263 &  3393 & 26.87 &  289 & 3093 & -0.315 & 0.761 & 0.042 \\
st28gm05n032 &  1.5 &  6936 & 637$^3$ & 1263 &  6295 & 24.93 &  280 & 3149 & -0.286 & 0.701 & 0.052 \\
st28gm05n033 &  1.5 &  6702 & 559$^3$ & 3454 &  6994 & 27.70 &  304 & 2993 & -0.358 & 0.812 & 0.049 \\
st28gm05n034 &  1.5 &  6840 & 637$^3$ & 1263 &  7619 & 30.18 &  279 & 3141 & -0.285 & 0.709 & 0.038 \\
st28gm06n031 &  1.0 &  7016 & 281$^3$ & 1381 &  3191 & 25.27 &  362 & 2774 & -0.684 & 1.407 & 0.256 \\
st28gm06n032 &  1.0 &  7021 & 401$^3$ & 1970 &  3393 & 26.87 &  341 & 2861 & -0.631 & 1.391 & 0.190 \\
st28gm06n033 &  1.0 &  7033 & 281$^3$ & 1381 &  3392 & 26.86 &  365 & 2766 & -0.691 & 1.322 & 0.220 \\
st28gm06n037 &  1.5 & 10000 & 317$^3$ & 1558 &  3694 & 29.26 &  389 & 2926 & -0.569 & 1.164 & 0.127 \\
st28gm06n038 &  1.0 &  7049 & 401$^3$ & 1970 &  3401 & 26.94 &  363 & 2774 & -0.687 & 1.459 & 0.218 \\
st28gm06n039 &  1.0 &  7027 & 401$^3$ & 1970 &  3406 & 26.98 &  351 & 2821 & -0.656 & 1.275 & 0.271 \\
st28gm06n043 &  1.0 &  7039 & 317$^3$ & 1558 &  5405 & 42.81 &  346 & 2842 & -0.644 & 1.262 & 0.206 \\
st28gm06n045 &  1.0 &  7035 & 499$^3$ & 3365 &  3392 & 26.86 &  365 & 2767 & -0.690 & 1.526 & 0.237 \\
st28gm06n046 &  1.0 &  7022 & 499$^3$ & 3365 &  3401 & 26.94 &  363 & 2772 & -0.686 & 1.550 & 0.260 \\
st28gm06n050 &  1.0 &  7049 & 599$^3$ & 4858 &  6894 & 54.61 &  351 & 2823 & -0.656 & 1.396 & 0.205 \\
st28gm06n051 &  1.0 &  6933 & 637$^3$ & 1558 &  7618 & 48.27 &  328 & 2907 & -0.598 & 1.256 & 0.241 \\
st28gm06n052 &  1.0 &  7030 & 679$^3$ & 6386 &  3352 & 26.55 &  355 & 2806 & -0.665 & 1.493 & 0.243 \\
st28gm06n053 &  1.5 & 10016 & 317$^3$ & 1558 &  5639 & 35.73 &  393 & 2912 & -0.578 & 1.146 & 0.125 \\
 st28gm06n13 &  1.0 &  6932 & 281$^3$ & 1381 &  3783 & 29.96 &  352 & 2806 & -0.659 & 1.321 & 0.271 \\
 st28gm06n24 &  1.0 &  6944 & 281$^3$ & 1381 &  3001 & 23.77 &  341 & 2853 & -0.632 & 1.363 & 0.186 \\
 st28gm06n25 &  1.0 &  6890 & 401$^3$ & 1970 &  3001 & 23.77 &  340 & 2851 & -0.630 & 1.544 & 0.259 \\
 st28gm06n26 &  1.0 &  6955 & 281$^3$ & 1381 &  3201 & 25.35 &  342 & 2848 & -0.635 & 1.330 & 0.250 \\
 st28gm06n27 &  1.0 &  6950 & 281$^3$ & 1381 &  3201 & 25.35 &  341 & 2854 & -0.631 & 1.254 & 0.188 \\
 st28gm06n28 &  1.0 &  6969 & 281$^3$ & 1381 &  3201 & 25.35 &  340 & 2860 & -0.629 & 1.191 & 0.183 \\
st28gm07n002 &  1.0 &  9986 & 317$^3$ & 2222 &  5984 & 47.40 &  452 & 2714 & -0.875 & 1.852 & 0.552 \\
st28gm07n004 &  1.0 &  9958 & 397$^3$ & 2782 &  7379 & 58.45 &  445 & 2734 & -0.861 & 1.705 & 0.388 \\
st29gm04n001 &  1.0 &  4982 & 281$^3$ & 1381 &  3201 & 25.35 &  285 & 2871 & -0.478 & 0.929 & 0.103 \\
st29gm06n001 &  1.0 &  6948 & 281$^3$ & 1381 &  3201 & 25.35 &  328 & 2911 & -0.597 & 1.171 & 0.236 \\ \hline

\end{tabular}
\end{table*}
\clearpage \newpage
\addtocounter{table}{-1} 
     \begin{table*}[t]
         \caption{continued.}
         \centering
         \begin{tabular}{lcccccccccccc}
             \hline \hline
Model & $M_\star$ & $L_\star$ & $n_x^3$ & $x_\mathrm{box}$ & $n_t$ & $t_\mathrm{avg}$ & $R_{\star,s_\mathrm{min}}$ & $T_{\mathrm{eff},s_\mathrm{min}}$ & $\log g_{s_\mathrm{min}}$ & $P_\mathrm{puls}$ & $\sigma_\mathrm{puls}$ \\
 & $M_\sun$ & $L_\sun$ &  & $R_\sun$ &  & yr & $R_\sun$ & K & (cgs) & yr & yr \\
\hline
st34gm02n001 &  8.0 & 41017 & 637$^3$ & 1929 &  6673 & 16.91 &  615 & 3311 & -0.241 & 0.873 & 0.087 \\
st34gm02n002 &  8.0 & 41675 & 765$^3$ & 1626 &  9140 & 23.17 &  602 & 3359 & -0.223 & 0.859 & 0.088 \\
st34gm02n003 &  8.0 & 40794 & 765$^3$ & 1626 & 11402 & 28.90 &  600 & 3349 & -0.219 & 0.896 & 0.108 \\
st34gm03n001 &  5.0 & 28710 & 637$^3$ & 1929 &  7252 & 22.98 &  500 & 3358 & -0.266 & 0.869 & 0.111 \\
st34gm03n002 &  6.0 & 41195 & 637$^3$ & 1929 &  5761 & 18.25 &  611 & 3326 & -0.360 & 1.064 & 0.129 \\
st35gm03n020 & 12.0 & 89138 & 637$^3$ & 2093 &  6982 & 22.12 &  759 & 3620 & -0.246 & 0.945 & 0.077 \\
 st35gm03n09 & 12.0 & 83405 & 235$^3$ & 2027 &   489 &  7.73 &  836 & 3391 & -0.342 & 1.056 & 0.146 \\
 st35gm03n10 & 12.0 & 74135 & 235$^3$ & 2027 &   398 &  6.31 &  768 & 3436 & -0.263 & 0.780 & 0.108 \\
 st35gm03n11 & 12.0 & 72676 & 255$^3$ & 1751 &   940 & 14.89 &  752 & 3456 & -0.243 & 0.807 & 0.114 \\
 st35gm03n18 & 12.0 & 89067 & 235$^3$ & 2027 &   900 & 14.24 &  818 & 3487 & -0.321 & 0.940 & 0.132 \\
 st35gm03n19 & 12.0 & 88938 & 401$^3$ & 2027 &   600 &  4.75 &  772 & 3587 & -0.267 & 1.014 & 0.115 \\
st35gm04n045 &  5.0 & 41336 & 637$^3$ & 1929 &  6600 & 20.91 &  597 & 3368 & -0.419 & 1.412 & 0.160 \\
st35gm04n046 &  5.0 & 40856 & 637$^3$ & 1929 &  7391 & 23.42 &  600 & 3350 & -0.423 & 1.327 & 0.185 \\
st35gm04n047 &  5.0 & 40923 & 637$^3$ & 1929 &  8391 & 26.59 &  622 & 3291 & -0.455 & 1.426 & 0.166 \\
st35gm04n048 &  5.0 & 40914 & 637$^3$ & 1929 &  8390 & 26.58 &  623 & 3288 & -0.456 & 1.433 & 0.253 \\
st35gm04n049 &  5.0 & 40853 & 637$^3$ & 1929 &  9397 & 29.77 &  602 & 3344 & -0.426 & 1.398 & 0.169 \\
st35gm04n050 &  5.0 & 45416 & 637$^3$ & 1929 &  9591 & 30.39 &  611 & 3408 & -0.439 & 1.385 & 0.181 \\
st35gm04n051 &  5.0 & 40738 & 637$^3$ & 1929 &  9393 & 29.76 &  597 & 3355 & -0.419 & 1.364 & 0.260 \\
 st35gm04n26 &  5.0 & 41490 & 171$^3$ & 1626 &   977 & 15.46 &  657 & 3212 & -0.502 & 1.396 & 0.146 \\
 st35gm04n34 &  5.0 & 41815 & 315$^3$ & 1626 &   205 &  3.23 &  615 & 3328 & -0.445 & 1.099 & 0.257 \\
 st35gm04n36 &  5.0 & 41523 & 235$^3$ & 1626 &   901 & 14.26 &  620 & 3309 & -0.451 & 1.530 & 0.186 \\
 st35gm04n37 &  5.0 & 41227 & 315$^3$ & 1626 &   643 &  5.09 &  597 & 3366 & -0.419 & 1.159 & 0.172 \\
  st36g00n04 &  6.0 & 24216 & 255$^3$ & 1000 &  1983 &  6.28 &  388 & 3652 &  0.017 & 0.422 & 0.043 \\
\hline
\end{tabular}
\tablefoot{The table presents the model name, which is made up of the approximate effective temperature and surface gravity, and of a running number; the current stellar mass $M_\star$, used for the external potential; the average emitted luminosity $L_\star$, approximately identical to the inserted luminosity in the core; the model dimensions $n_x^{3}$; the edge length of the cubical computational box $x_\textrm{box}$; the temporal dimension $n_t$ and the total time $t_\textrm{avg}$, used for the averaging of the rest of the quantities in this table and for the further analysis; the average approximate stellar radius $R_\star$ from the entropy minimum definition, used to derive the following quantities; the average approximate effective temperature $T_\textrm{eff}$; the logarithm of the average approximate surface gravity $\log g$; the pulsation period $P_\textrm{puls}$; and the spread in the pulsation period $\sigma_\textrm{puls}$.}
\end{table*}

\end{appendix}

\end{document}